\documentclass[12pt]{article}
\usepackage[a4paper, left=2cm, top=2cm, right=2cm, bottom=1.5cm, includefoot]{geometry}

\usepackage{amsmath}
\usepackage{amsthm}
\usepackage{amssymb}
\usepackage{bm}
\usepackage{tabularx}
\usepackage{graphicx}
\usepackage{arydshln}
\usepackage{subfigure}
\usepackage{color}
\usepackage[pdftex]{hyperref}
\usepackage{listings}
\usepackage{array}

\newcolumntype{C}[1]{>{\centering\let\newline\\\arraybackslash\hspace{0pt}}m{#1}}

\hypersetup{%
  colorlinks=true,
  urlcolor=blue,
  citecolor=black
}

\title{Compact difference scheme for parabolic and Schr\"odinger-type equations with variable coefficients}

\author{Vladimir A. Gordin 
    \thanks{National Research University Higher School of Economics,
    Hydrometeorological Center of Russia, Moscow 123242, Russia. Email: vagordin@mail.ru (corresponding author)}
\and Evgenii A. Tsymbalov \thanks{National Research University Higher School of Economics, 
    Skolkovo Institute for Science and Technology, Russia. Email: etsymbalov@gmail.com}}

\date{}

\begin{document}

\maketitle 

\begin{abstract}

    We develop a new compact scheme for the second-order PDE (parabolic and Schr\"odinger type)
    with a variable time-independent coefficient. It has a higher order and smaller error than classic implicit scheme. The Dirichlet and Neumann boundary problems are considered. The relative finite-difference operator is almost self-adjoint.
    
\end{abstract}

\section{Introduction}

\subsection{Stationary problems}

The most popular implicit finite difference schemes, which approximate classic PDEs of 
mathematical physics, use three point stencils (for 1D problems) and have the second order of approximation. To improve the order, we can develop the stencil up to
five points, however, in this case there are two significant obstacles: 
\begin{itemize}
    \item some additional boundary conditions are needed in comparison with the corresponding
differential boundary problem;
    \item a linear algebraic system that we solve at every temporal step has to be solved with a five-diagonal matrix instead of
a three-diagonal one, and therefore the number of arithmetic operations is doubled compared to 
the corresponding computational implementation of such a scheme.
\end{itemize}

There is an alternative approach to improve the order: to use compact 
finite difference schemes. We can optimally average the right-hand side of the corresponding original differential equation. For instance, we can approximate the ordinary differential equation
\begin{equation}
\label{diffusion0}
    d^2_xu=f, \;x \in [a,b],\; u(a) = A,\: u(b) = B,
\end{equation}
by the compact finite difference equation on the equidistant grid ${\{ x_j \}}^N_{j=0},\;
 x_j=a+jh,\; h=(b-a)/N$:
\begin{equation}\label{cscheme}
    u_{j-1} - 2u_j + u_{j+1} = h^2[f_{j-1} + 10f_j + f_{j+1}]/12,\quad j=1,\ldots, N-1
\end{equation}
instead of the classic finite difference equation
\[
    u_{j-1} - 2u_j + u_{j+1} = h^2f_{j}.
\]

Here $f_j = f(x_j)$ is a given function; $u_j$ is an approximation of the
unknown solution $u(x), u_0=A, u_N=B$. The double-sweep method can be used to invert the same three-diagonal matrix and to obtain the solution 
${\{ u_j \}}^{N-1}_{j=1}$ with a better approximation \cite{gord-10}.

Similarly, we can use the scheme
\begin{equation*}
\begin{aligned}
    u_{i, j} + 0.2(u_{i, j - 1} + &u_{i - 1, j} + u_{i, j + 1} + u_{i + 1, j}) - 0.05(u_{i - 1, j - 1} + u_{i - 1, j + 1} + u_{i + 1, j - 1} + u_{i + 1, j + 1}) = 
    \\
    &-0.2h^2f_{i, j} - 0.025h^2(f_{i, j - 1} + f_{i - 1, j} + f_{i, j + 1} + f_{i + 1, j}),
\end{aligned}
\end{equation*}
to approximate with the $4$-th order the Poisson equation \begin{equation*}
    \Delta u = f(x,y),
\end{equation*}
on a rectangular equidistant grid instead of the second order classic implicit scheme:
\begin{equation*}
    u_{i, j - 1} + u_{i - 1, j} + u_{i, j + 1} + u_{i + 1, j} - 4u_{i, j} = h^2f_{i, j} .
\end{equation*}

Similar compact high order schemes were used after the fast Fourier transform with respect to 
longitude in \cite{gord-10} for the solution of an elliptic system of PDEs on $\mathbb{S}^2$. The 
equations have singularities at the points of the poles, and special boundary conditions at the ends of 
the segment $[-\pi/2,\, \pi/2]$ (according to \cite{abramov71}) with respect to 
latitude. There is a separate asymptotic in the poles for every Fourier-mode. The computational 
effectiveness is essential here because this elliptic system is an important ingredient of weather 
forecasting models. It is applied to the every vertical level on every time step in the forecasting 
model, see e.g. \cite{tolstykh02}.

There are a few ways to determine coefficients for the compact difference scheme for a given PDE. One of the main ideas is to use a truncated Taylor series expansion (\cite{gupta1984single, zhang1998explicit, gelu2017sixth}); in \cite{sutmann2007compact} it is used together with the Pade approximation. Another approach is to utilize the central difference by expanding the leading truncation error term until the desirable order is reached (\cite{spotz1996high, ge2002symbolic}). For both approaches symbolic computations are used extensively to get rid of exhaustive algebraic manipulations. In our works (\cite{gord-14, gordin-05}), we develop a much simpler approach based on undetermined coefficients, which also uses computational algebra packages to derive the coefficients of the compact scheme. However, the majority of the formulae have been constructed for linear differential equations with 
constant coefficients only. 
                                            
There was an exception: differential equations with a variable coefficient in the low-order term. For 
instance, the ordinary differential equation
\begin{equation*}
    d^2_xu + \rho(x)u = f(x)
\end{equation*}
may be approximated in the following way:
\begin{equation*}
\begin{aligned}
    &u_{j-1} -2u_j + u_{j+1} = h^2{[f_{j-1} - \rho(x_{j-1})u_{j-1} ] + 10[f_j - \rho(x_j)u_j] + [f_{j+1} - \rho(x_{j+1})u_{j+1}]}/12 \implies
    \\
    &[1 + h^2\rho(x_{j-1})/12]u_{j-1} + 2[5h^2\rho(x_j)/12-1]u_j + [1 + h^2\rho(x_{j+1})/12]u_{j+1} = 
    \\
    &h^2[f_{j-1} + 10f_j + f_{j+1}]/12,
\end{aligned}
\end{equation*}
--- it is a corollary of the relation (\ref{cscheme}).

If the coefficient $\rho(x)$ is a non-positive function, then the corresponding 
three-diagonal matrix is (according to the Gershgorin theorem, see e.g. \cite{gord-10}) 
negative definite. 

\subsection{Evolutionary PDEs with a single spatial variable}

The computational approach, which uses compact high order schemes can be developed for 
evolutionary PDEs, e.g. for the diffusion equation and for the Schr\"odinger one, see e.g. 
\cite{gord-14}.

Compact difference schemes can be also developed for linear PDEs with a variable coefficient 
in the low-order term, e.g. to diffusion equation or Schr\"odinger equation with a potential.
In this work, we focus on the important type of PDEs: 1D parabolic equations. Namely,
we approximate the diffusion equation with a variable smooth positive coefficient: 
\begin{equation} \label{diffusion_eq}
 \frac{\partial u}{\partial t}=Pu+f,\quad Pu=
\frac{\partial}{\partial x}\theta(x) \frac{\partial u}{\partial x},
\end{equation}
where $\theta(x): \mathbb{R} \rightarrow \mathbb{R}_+$ is a variable time-independent 
diffusion coefficient, $t \in [0; T],x \in [0, 2\pi]$, $f=f(t, x)$ is a forcing. 

Then we consider the Leontovich -- Levin (Schr\"odinger-type) equation 
$i \frac{\partial u}{\partial t}=Pu+f $, and construct a similar compact scheme for it.

Earlier we constructed the $4$-th order compact finite difference scheme for the first 
boundary problem for the linear ordinary self-adjoint equation:
\begin{equation}\label{sturm}
    - d_x[\theta(x)d_xu] = f(x), ~ x \in [a, b],
\end{equation}
where the coefficient $\theta(x)$ is strongly positive and smooth \cite{gt16b}. If the coefficient $\theta(x)$ is discontinuous at the point $x^* \in (a, b)$ in equation (\ref{sturm}), the special confinement boundary conditions are necessary to provide fulfillment of the mass (or energy) conservation law as well as the high convergence rate, see \cite{gt16a}.

\textbf{Note.} The linear operator $P$ is self-adjoint in the space of smooth functions under homogeneous Dirichlet conditions in the sense of Hilbert metric $L^2 [0, 2\pi]$. The spectrum of the self-adjoint operator $P$ is real
and negative. Therefore, the resolving operators $R(t) = \exp(Pt), t > 0$
of the mixed initial-boundary problem in the space is self-adjoint and contractive.

\textbf{Note.} The linear operator $P$ under periodical or homogeneous Neumann boundary conditions is $L^2$-self-adjoint, too. However it is non-positive, and resolving operators $R(t)$ are contractive on the orthogonal complement to the one-dimensional subspace of constants in $L^2$.
The case of the Neumann boundary conditions is more difficult than the Dirichlet one for the compact finite-difference approximation. To provide high order of the scheme, we use wide stencils at the endpoints in two time moments. If the simplest approximation $u_0 = u_1 = 0$ of the Neumann conditions is used, we obtain the first order of error decrease instead of the fourth one.

\subsection{Multidimensional problems}

Multidimensional equation
\begin{equation}  \label{diff_multi}
\rho\partial_t u= {\bf div}\,\vartheta \,  {\bf grad}\, u +f,
\end{equation}
where $\vec x =\langle x_1,\,\ldots,\,x_n\rangle \in G\subseteq \mathbb{R}^n,\;\rho=\rho(\vec x),\:\vartheta =\vartheta(\vec x),\:f=f(t,\,\vec x)$ is a natural generalization of the equation~(\ref{diffusion_eq}). 

If the coefficients $\rho$ and $\vartheta$ are constant, then the coefficients of the relative compact scheme are obtained without strong difficulties. Certainly, the result depends on the choice of the difference scheme grid: rectangular, triangular, or hexagonal. The compact scheme for the Poisson equation may be constructed for such grids as well as the rectangular grid in polar coordinates, see \cite{gordin-05, lai2007formally}. The compact schemes may be developed (\cite{gord-14, tangman2008numerical, karaa2009two}) for diffusion equation with a constant coefficient, too.

The case of equation (\ref{diff_multi}) with an arbitrary smooth variable coefficient $\theta$ is more difficult. However, the case of the coefficient, which depends on one variable $x_1$ only, may be reduced to the case considered here if the area $G$ is a direct sum of segments and/or circumferences, i.e. if we can use the fast Fourier transform with respect to variables $x_2,\,\ldots,\,x_n$. Such examples for the Helmholtz equation (or similar system) was considered in \cite{gord-10, tolstykh02} for $G=\mathbb{S}^2$, where $x_1$ is the latitude, $x_2$ is the longitude, see also \cite{chen2011optimal}. The special kind of boundary conditions (individual for any longitude Fourier mode) should be used here at the polar points. Otherwise, we do not obtain a desirable order of approximation.  The compact schemes for the 4-th order of approximation of the second order elliptical linear PDE (Helmholtz-type) with a variable coefficient were constructed in \cite{ge2002symbolic, britt2011numerical}. The fourth order elliptical PDE with variable coefficient was considered in \cite{wang2008fourth}.

The rest of the paper is organized as follows. In Sect. 2 we describe the "compact approach" to finite-difference approximation: for the positive smooth coefficient $\theta$ (Sect. 2.1), for inner grid points (Sect. 2.2), for the Neumann boundary conditions (Sect. 2.3). We also introduce the classic second order implicit scheme (Sect. 2.4) and the Leontovich -- Levin equation (Sect. 2.5).

In Sect. 3, which is devoted to numerical experiments, we introduce sample solutions for numerical experiments (Sect. 3.1), examine approximation order numerically in Sect. 3.2 and utilize the Richardson extrapolation approach to finite-difference scheme's improvement (Sect. 3.3). The possible simplification of the scheme's coefficients is tested numerically in Sect. 3.4; the spectra of transition operators are analyzed in Sect. 3.5. Similar constructions for the Leontovich -- Levin equation are examined in Sect. 3.6. 

Section 4 concludes the paper.


\section{Compact difference scheme}

\subsection{Diffusion coefficient approximation for the $4-$th order finite difference model}
\begin{figure}[h!]
\begin{center}
	\includegraphics[scale=1]{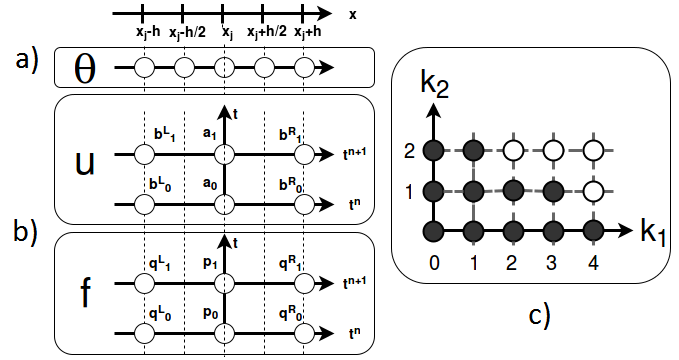}
	\caption{
    \textbf{a, b}: stencils for the compact finite difference scheme. \textbf{c}: diagram for test functions $u^*_{k_1, k_2} = y^k_1 t^k_2$, which are used in order to obtain the coefficients of scheme (\ref{comp_scheme}). Monomials, which are denoted as white circles, are unnecessary to obtain the coefficients yet the equation (\ref{comp_scheme}) holds on them.
	}
    \label{fig:stencil}
\end{center}
\end{figure}

The coefficient $\theta(x)$ in equation (\ref{diffusion_eq}) for all the physical problems is non-negative. Otherwise, the Cauchy problem for equation (\ref{diffusion_eq}) is incorrect in the Hadamard sense. The special case when $\theta(x)$ is non-negative and has zeros is not considered here.
For the compact scheme construction we need to explicitly determine the derivatives of $\theta(x)$ in the grid points. Since the coefficient $\theta(x)$ is smooth and strongly positive, we approximate it 
locally (in the vicinity of an arbitrary internal grid point $x_j$) by the exponential function $\theta(x) = exp(\rho(x))$, where $\rho$ is an arbitrary real function. To ensure that the resulting compact scheme has a high order, we approximate $\rho(x)$ by the $4$-th order polynomial:
\begin{equation} \label{exp_approx}
 \theta(x) \approx \theta(x_j) exp(c_1 y + c_2 y^2 + c_3 y^3 + c_4 y^4),
\end{equation}
where $y = x - x_j$, and then we determine the coefficients $c_1, c_2, c_3, c_4$ by using the interpolation 
conditions. We assume that relation (\ref{exp_approx}) is exact at the points 
$y = -h, -h/2, h/2, h$, see Fig.~\ref{fig:stencil}a. Thus, for every $j$ we obtain four linear algebraic equations, where $\theta_k=\theta(x_k)$, for these four undetermined coefficients:

$c_4 h^4 - c_3 h^3 + c_2 h^2 - c_1 h = \ln\left(\theta_{j-1}/\theta_j\right)$,

$c_4 h^4/16 - c_3 h^3/8 + c_2 h^2/4 - c_1 h/2 = \ln\left(\theta_{j-\frac{1}{2}}/\theta_j\right)$,

$c_4 h^4/16 + c_3 h^3/8 + c_2 h^2/4 + c_1 h/2 = \ln\left(\theta_{j+\frac{1}{2}}/\theta_j\right)$,

$c_4 h^4 + c_3 h^3 + c_2 h^2 + c_1 h = \ln\left(\theta_{j+1}/\theta_j\right)$.

We solve the system and obtain the following solution $c_1, c_2, c_3, c_4$, where $\theta_k=\theta(x_k)$:


$ c_1 = -[8\,  \ln\left(\theta_{j-\frac{1}{2}}/\theta_j\right) - 8\, \ln\left(\theta_{j+\frac{1}{2}}/\theta_j\right) - \ln\left(\theta_{j-1}/\theta_j\right) + \ln\left(\theta_{j+1}/\theta_j\right)]/6h$,

$ c_2 = [16\, \ln\left(\theta_{j-\frac{1}{2}}/\theta_j\right) + 16\, \ln\left(\theta_{j+\frac{1}{2}}/\theta_j\right) - \ln\left(\theta_{j-1}/\theta_j\right) - \ln\left(\theta_{j+1}/\theta_j\right)]/6h^2$,

$ c_3 = 2\,[ 2\, \ln\left(\theta_{j-\frac{1}{2}}/\theta_j\right) - 2\, \ln\left(\theta_{j+\frac{1}{2}}/\theta_j\right) - \ln\left(\theta_{j-1}/\theta_j\right) +  \ln\left(\theta_{j+1}/\theta_j\right)]/3h^3$,

$ c_4 = -2\,[4\, \ln\left(\theta_{j-\frac{1}{2}}/\theta_j\right) + 4\, \ln\left(\theta_{j+\frac{1}{2}}/\theta_j\right) - \ln\left(\theta_{j-1}/\theta_j\right) -  \ln\left(\theta_{j+1}/\theta_j\right)]/3h^4$.

In the simplest case of $\theta(x)=$~const we obtain the trivial solution: $c_1 = c_2 = c_3 = c_4 = 0$.

\subsection{Test functions and the corresponding coefficients for the implicit compact 
scheme}

We construct the scheme on six-point two-layer stencils (see Fig. \ref{fig:stencil}b):
\begin{equation}  \label{comp_scheme}
\begin{aligned}
&b^L_{0,j}u^n_{j-1} + a_{0,j}u^n_j + b^R_{0,j}u^n_{j+1} + b^L_{1,j}u^{n+1}_{j-1} + a_{1,j}u^{n+1}_j + b^R_{1,j}u^{n+1}_{j+1} = 
\\
&q^L_{0,j}f^n_{j-1} + p_{0,j}f^n_j + q^R_{0,j}f^n_{j+1} + q^L_{1,j}f^{n+1}_{j-1} + p_{1,j}f^{n+1}_j + q^R_{1,j}f^{n+1}_{j+1},    
\end{aligned}
\end{equation}
$ j = 1 \ldots N-1, \: n = 0 \ldots T.$ We assume that the relation (\ref{comp_scheme}) holds for several test exact solutions of equation (\ref{diffusion_eq}): 
$\langle u^\star,\,f^\star \rangle$.

We use here the basis of test functions
\begin{equation}
\label{testest}
 u^*_{k_1,\,k_2} = y^{k_1}t^{k_2}, f^*_{k_1,\,k_2} = \frac{\partial  u^*_{k_1,\,k_2}}{\partial t} - P  u^*_{k_1,\,k_2},
\end{equation}
shown by black circles on the diagram $\langle
k_1,\,k_2\rangle$ (Fig.~\ref{fig:stencil}b). We substitute all of them to relation (\ref{comp_scheme}) to obtain a system of 11 homogeneous linear algebraic 
equations for the coefficients of compact scheme (\ref{comp_scheme}) at an arbitrary inner grid points $x_j, j \neq 0, N$.

We add to the system one normalizing linear non-homogeneous equation $a_{0,j} = C^* = const > 0$ (see \cite{gord-10, gord-14, gt16b, gt16a}) and solve the $12$-th linear algebraic system; see the obtained $12$ coefficients in Table ~\ref{tab:coeffsu}, ~\ref{tab:coeffsf}.

\begin{table}
\caption{Coefficients for the left-hand side of compact scheme (\ref{comp_scheme}), expanded with respect to degrees of $h$. Here, $r_- = \theta(x_j)/\theta(x_j-h),\; r_+ = \theta(x_j)/\theta(x_j+h)$. We choose the normalizing constant $C^*$ in a way to provide the coefficients in the most algebraically simple form.}
\label{tab:coeffsu}
\centering
\begin{tabular}{|c|m{2cm}|m{2cm}|m{2cm}|m{2cm}|m{2cm}|m{2cm}|} 
\hline
- & $a_{0,j}$ & $b^L_{0,j}$ & $b^R_{0,j}$ & $a_{1,j}$ & $b^L_{1,j}$ & $b^R_{1,j}$ \\  
\hline
$1$ & $144 \nu_j - 120$ & $- 72 \nu_j - 12 r_-$ & $- 72 \nu_j - 12 r_+$ & $144 \nu_j + 120$ & $12 r_- - 72 \nu_j$ & $- 72 \nu_j + 12 r_+$ \\  
\hline
$h$ & $0$ & $36 c_1 \nu_j + 6 c_1 r_-$ & $- 36 c_1 \nu_j - 6 c_1 r_+$ & $0$ & $36 c_1 \nu_j - 6 c_1 r_-$ & $6 c_1 r_+ - 36 c_1 \nu_j$ \\  
\hline
$h^2$ & $8 c_1^2 - 128 c_2 + 192 c_2 \nu_j$ & $2 r_- c_1^2 - 96 c_2 \nu_j - 8 c_2 r_-$ & $2 r_+ c_1^2 - 96 c_2 \nu_j - 8 c_2 r_+$ & $- 8 c_1^2 + 128 c_2 + 192 c_2 \nu_j$ & $- 2 r_- c_1^2 - 96 c_2 \nu_j + 8 c_2 r_-$ & $- 2 r_+ c_1^2 - 96 c_2 \nu_j + 8 c_2 r_+$ \\  
\hline
$h^3$ & $0$ & $18 c_3 \nu_j - 12 c_3 r_- - 3 c_1^3 \nu_j + 42 c_1 c_2 \nu_j + 4 c_1 c_2 r_-$ & $12 c_3 r_+ - 18 c_3 \nu_j + 3 c_1^3 \nu_j - 42 c_1 c_2 \nu_j - 4 c_1 c_2 r_+$ & $0$ & $18 c_3 \nu_j + 12 c_3 r_- - 3 c_1^3 \nu_j + 42 c_1 c_2 \nu_j - 4 c_1 c_2 r_-$ & $3 c_1^3 \nu_j - 12 c_3 r_+ - 18 c_3 \nu_j - 42 c_1 c_2 \nu_j + 4 c_1 c_2 r_+$  \\  
\hline
$h^4$ & $48 c_1 c_3 - 256 c_4 + 384 c_4 \nu_j + 64 c_2^2 \nu_j - 32 c_2^2 - 48 c_1 c_3 \nu_j$ & $- 32 \nu_j c_2^2 - 192 c_4 \nu_j - 16 c_4 r_- + 24 c_1 c_3 \nu_j + 6 c_1 c_3 r_-$ & $- 32 \nu_j c_2^2 - 192 c_4 \nu_j - 16 c_4 r_+ + 24 c_1 c_3 \nu_j + 6 c_1 c_3 r_+$ & $256 c_4 - 48 c_1 c_3 + 384 c_4 \nu_j + 64 c_2^2 \nu_j + 32 c_2^2 - 48 c_1 c_3 \nu_j$ & $- 32 \nu_j c_2^2 - 192 c_4 \nu_j + 16 c_4 r_- + 24 c_1 c_3 \nu_j - 6 c_1 c_3 r_-$ & $- 32 \nu_j c_2^2 - 192 c_4 \nu_j + 16 c_4 r_+ + 24 c_1 c_3 \nu_j - 6 c_1 c_3 r_+$  \\  
\hline
$h^5$ & $0$ & $84 c_1 c_4 \nu_j + 8 c_1 c_4 r_- + 12 c_1 c_2^2 \nu_j - 18 c_1^2 c_3 \nu_j$ & $18 c_1^2 c_3 \nu_j - 8 c_1 c_4 r_+ - 12 c_1 c_2^2 \nu_j - 84 c_1 c_4 \nu_j$ & $0$ & $84 c_1 c_4 \nu_j - 8 c_1 c_4 r_- + 12 c_1 c_2^2 \nu_j - 18 c_1^2 c_3 \nu_j$ & $8 c_1 c_4 r_+ - 84 c_1 c_4 \nu_j - 12 c_1 c_2^2 \nu_j + 18 c_1^2 c_3 \nu_j$ \\  
\hline
$h^6$ & $72 c_3^2 - 144 c_3^2 \nu_j - 128 c_2 c_4 + 256 c_2 c_4 \nu_j$ & $72 \nu_j c_3^2 - 128 c_2 c_4 \nu_j$ & $72 c_3^2 \nu_j - 128 c_2 c_4 \nu_j$ & $128 c_2 c_4 - 144 c_3^2 \nu_j - 72 c_3^2 + 256 c_2 c_4 \nu_j$ & $72 \nu_j c_3^2 - 128 c_2 c_4 \nu_j$ & $72 c_3^2 \nu_j - 128 c_2 c_4 \nu_j$  \\  
\hline
$h^7$ & $0$ & $- 27 c_1 \nu_j c_3^2 + 48 c_1 c_2 c_4 \nu_j$ & $27 c_1 c_3^2 \nu_j - 48 c_1 c_2 c_4 \nu_j$ & $0$ & $- 27 c_1 \nu_j c_3^2 + 48 c_1 c_2 c_4 \nu_j$ & $27 c_1 c_3^2 \nu_j - 48 c_1 c_2 c_4 \nu_j$ \\  
\hline
$h^8$ & $256 c_4^2 \nu_j - 128 c_4^2$ & $-128 c_4^2 \nu_j$ & $- 128 c_4^2 \nu_j$ & $256 c_4^2 \nu_j + 128 c_4^2$ & $-128 c_4^2 \nu_j$ & $- 128 c_4^2 \nu_j$  \\  
\hline
$h^9$ & $0$ & $48 c_1 c_4^2 \nu_j$ & $- 48 c_1 c_4^2 \nu_j$ & $0$ & $48 c_1 c_4^2 \nu_j$ & $- 48 c_1 c_4^2 \nu_j$ \\  
\hline
\end{tabular}
\end{table}

\begin{table}
\centering
\caption{Coefficients for the right-hand side of compact scheme (\ref{comp_scheme}), expanded with respect to degrees of $h$.}
\label{tab:coeffsf}
\begin{tabular}{|c|c|c|c|} \hline
- & $p_{0,j} = p_{1,j}$ & $q^L_{0,j} = q^L_{1,j}$ & $q^R_{0,j} = q^R_{1,j}$ \\  \hline
$1$ & $60$ & $6 r_-$ & $6 r_+$ \\  \hline
$h$ & $0$ & $- 3 c_1 r_-$ & $3 c_1 r_+$ \\  \hline
$h^2$ & $4 (- c_1^2 + 16 c_2)$ & $r_- (4 c_2 - c_1^2)$ & $r_+ (4 c_2 - c_1^2)$ \\  \hline
$h^3$ & $0$ & $r_- (6 c_3 - 2 c_1 c_2)$ & $- r_+ (6 c_3 - 2 c_1 c_2)$  \\  \hline
$h^4$ & $4 (4 c_2^2 + 32 c_4 – 6 c_1 c_3)$ & $r_- (8 c_4 - 3 c_1 c_3)$ & $r_+ (8 c_4 - 3 c_1 c_3)$  \\  \hline
$h^5$ & $0$ & $- 4 c_1 c_4 r_-$ & $4 c_1 c_4 r_+$ \\  \hline
$h^6$ & $4 (- 9 c_3^2 + 16 c_2 c_4)$ & $0$ & $0$  \\  \hline
$h^7$ & $0$ & $0$ & $0$ \\  \hline
$h^8$ & $64 c_4^2$ & $0$ & $0$ \\  \hline
\end{tabular}
\end{table}

{\bf Note}. If these 11 linear algebraic equations hold, the algebraic linear homogeneous connections, which correspond to the white
circles on Fig.~\ref{fig:stencil}, hold without any additional conditions. In other words,
the rank of the $15$-th order matrix of the homogeneous linear algebraic system, which corresponds to all test functions 
(\ref{testest}) at $k_1=0,\ldots,4,\;k_2=0,\,1,\,2$, is equal to $11$ only.

{\bf Note}. The finite difference compact scheme for the diffusion equation strongly corresponds to
the similar compact scheme for the ordinary differential equation $-d_x \theta (x)d_x u=f$.
Let us substitute $u^n_j$ and $u^{n+1}_j$ with $u^*_j$ in the equation (\ref{comp_scheme}).
In other words, we calculate the coefficients 
$a_j=a_{0,j}+a_{1,j}$,  $b^L_j=b^L_{0,j}+b^L_{1,j}$, and  $b^R_j=b^R_{0,j}+b^R_{1,j}$.
We obtain the ordinary finite difference equation 
\[
b^L_j u_{j-1}^* + a_j u_{j}^*+b^R_j u_{j+1}^* =
2\left[
q^L_{j}f_{j-1} + p_{j}f_j + q^R_{j}f_{j+1}\right],\quad j=1,\,\ldots,\,N-1.
\]

If we divide the equation by the function $\nu_j = \theta(x_j)\tau h^{-2}$, we obtain exactly the $4$-th order
compact scheme, which approximates a linear $2$-nd order ordinary differential equation, see \cite{gt16a}.
It is not a trivial statement, because the coefficients of compact scheme (\ref{comp_scheme})
can be changed, if we use another set of test functions instead of set (\ref{testest}).
\subsection{Compact approximation of the Neumann boundary conditions}
Let us consider the equation~(\ref{diffusion_eq}) under the Neumann boundary conditions
\begin{equation*}
\frac{\partial u}{\partial x}\Bigr|_{x=0, L} = 0.
\end{equation*}
We consider here the compact approximation for the left boundary in the form:
\begin{equation} \label{neumann_coef}
    \sum\limits_{j=0}^2 \alpha^1_j u^{n+1}_{j h} + \sum\limits_{j=0}^2 \alpha^0_j u^{n}_{j h} = 
    \sum\limits_{j=0}^2 \beta^1_j f^{n+1}_{j h} + \sum\limits_{j=0}^2 \beta^0_j f^{n}_{j h},
\end{equation}
where $\alpha^k_j, \beta^k_j$, $j = 0,1,2; k = 0, 1$ are the coefficients which are determined by the basis of test functions $<u^{**}_{k_1, k_2}, f^{**}_{k_1, k_2}>$: $u^{**}_{k_1, k_2} \in \{ 1, t, t^2, x^2, x^2 t, x^2 t^2, x^3, x^3 t, x^3 t^2, x^4\}$; $f^{**}_{k_1, k_2} = \frac{\partial  u^{**}_{k_1, k_2}}{\partial t} - P  u^{**}_{k_1, k_2}$. This set of basis functions is a subset of one on the diagram $\langle k_1, k_2 \rangle$ on Fig.~\ref{fig:stencil}. 

We do not use here the test functions $u^{**}_{k_1, k_2} \in \{x, x t, x t^2\}$ as they are in contrary to the boundary conditions, e.g. if $u^{**} = x $, then $\frac{\partial u^{**}}{\partial x}\Bigr|_{x=0} = 1 \neq 0$. 

We construct the linear algebraic system for the coefficients of equation~(\ref{neumann_coef}) and obtain the following solution: 

\begin{itemize}
    \item $\alpha^1_0 = 6 \nu_0 + 4 a h + 8 b h^2 + 12 c h^3 + 16 d h^4 + 17 a h \nu_0 + 34 b h^2 \nu_0 + 51 c h^3 \nu_0 + 68 d h^4 \nu_0 + 8;$ 
    \item $\alpha^1_1 = 16 exp(-a h -b h^2 -c h^3 -d h^4) - 32 b h^3 \nu_0 - 48 c h^3 \nu_0 - 64 d h^4 \nu_0 - 16 a h \nu_0;$
    \item $\alpha^1_2 = - \nu_0 (4 d h^4 + 3 c h^3 + 2 b h^2 + a h + 6);$
    \item $\alpha^0_0 = 6 \nu_0 - 4 a h - 8 b h^2 - 12 c h^3 - 16 d h^4 + 17 a h \nu_0 + 34 b h^2 \nu_0 + 51 c h^3 \nu_0 + 68 d h^4 \nu_0 - 8;$
    \item $\alpha^0_1 = - 16 a h \nu_0 - 32 b h^2 \nu_0 - 48 c h^3 \nu_0 - 64 d h^4 \nu_0 - 16 exp(-a h -b h^2 -c h^3 -d h^4);$
    \item $\alpha^0_2 = - \nu_0 (4 d h^4 + 3 c h^3 + 2 b h^2 + a h + 6);$ 
    \item $\beta^0_0 = \beta^1_0 = 2 \tau \nu_0 (4 d h^4 + 3 c h^3 + 2 b h^2 + a h + 2);$
    \item $\beta^0_1 = \beta^1_1 = 8 \tau \nu_0  exp(- d h^4 - c h^3 - b h^2 - a h);$
    \item $\beta^0_2 = \beta^1_2 = 0.$
\end{itemize}

We also obtain the similar coefficients for the Neumann boundary conditions at the right boundary. We note that $4-th$ approximation order cannot be obtained on a two-point stencil for boundary conditions as the linear algebraic system will be over-determined because we obtain too many equations for the selected number of variables (coefficients). 

\begin{figure}[h!]
    \centering
    \begin{minipage}{0.45\textwidth}
        \centering
        \includegraphics[scale=.4]{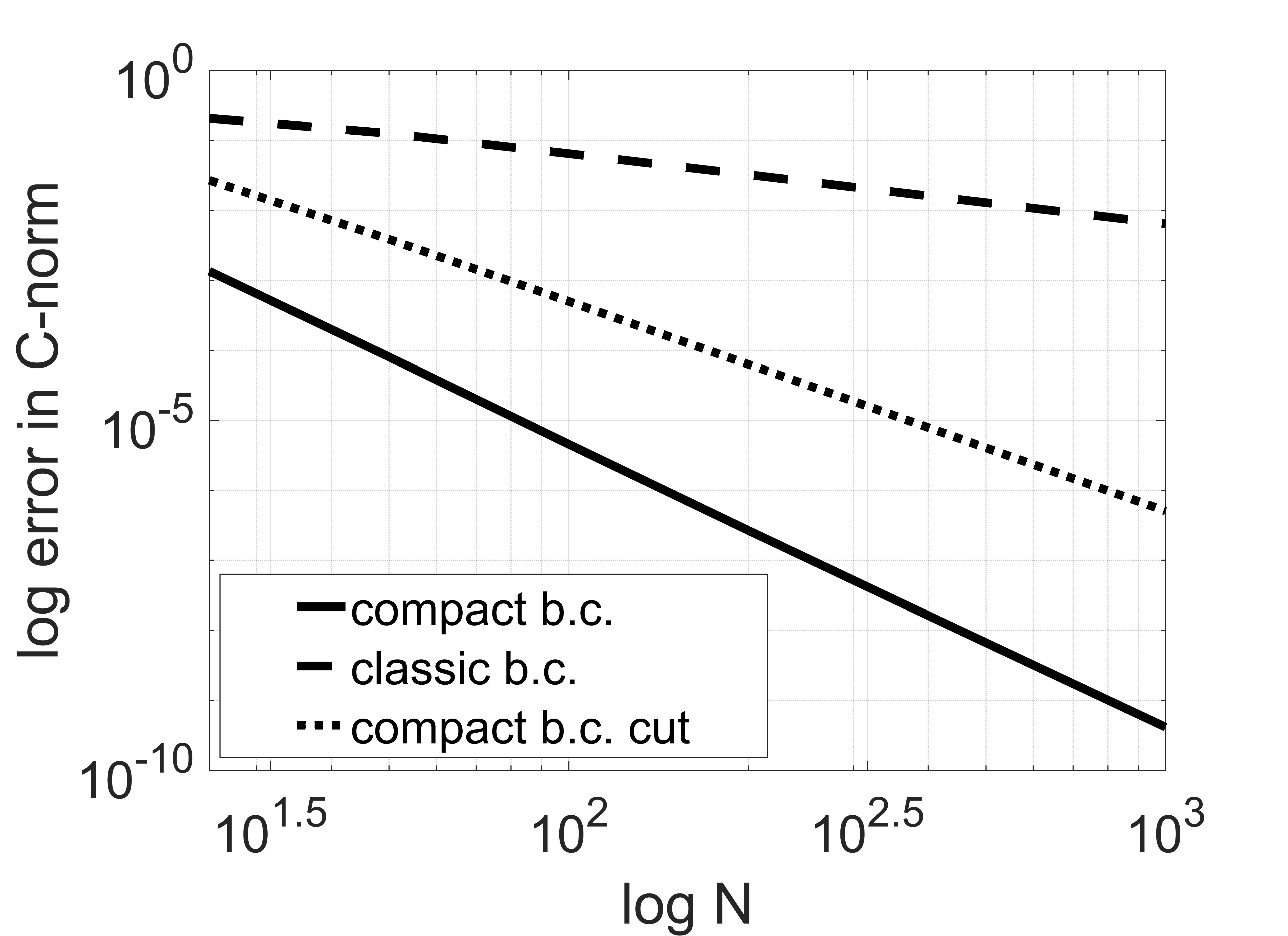}
    	\caption{
    Errors of the compact scheme on sample solutions (\ref{test_neumann}) of diffusion equation (\ref{diffusion_eq}) at $\nu^{\star} = 5,\: T = 1$. Bilogarithmic scale. Joint usage of compact difference scheme (\ref{ll_comp_scheme}) and boundary conditions approximation (\ref{neumann_coef}) shows the $4^{th}$ error decrease rate, while classic approximation (with $\epsilon = 0.5$) decreases the error rate down to $1^{st}$.
    	}
	\label{fig:neumann}
    \end{minipage}\hfill
    \begin{minipage}{0.45\textwidth}
        \centering
        \includegraphics[scale=.32]{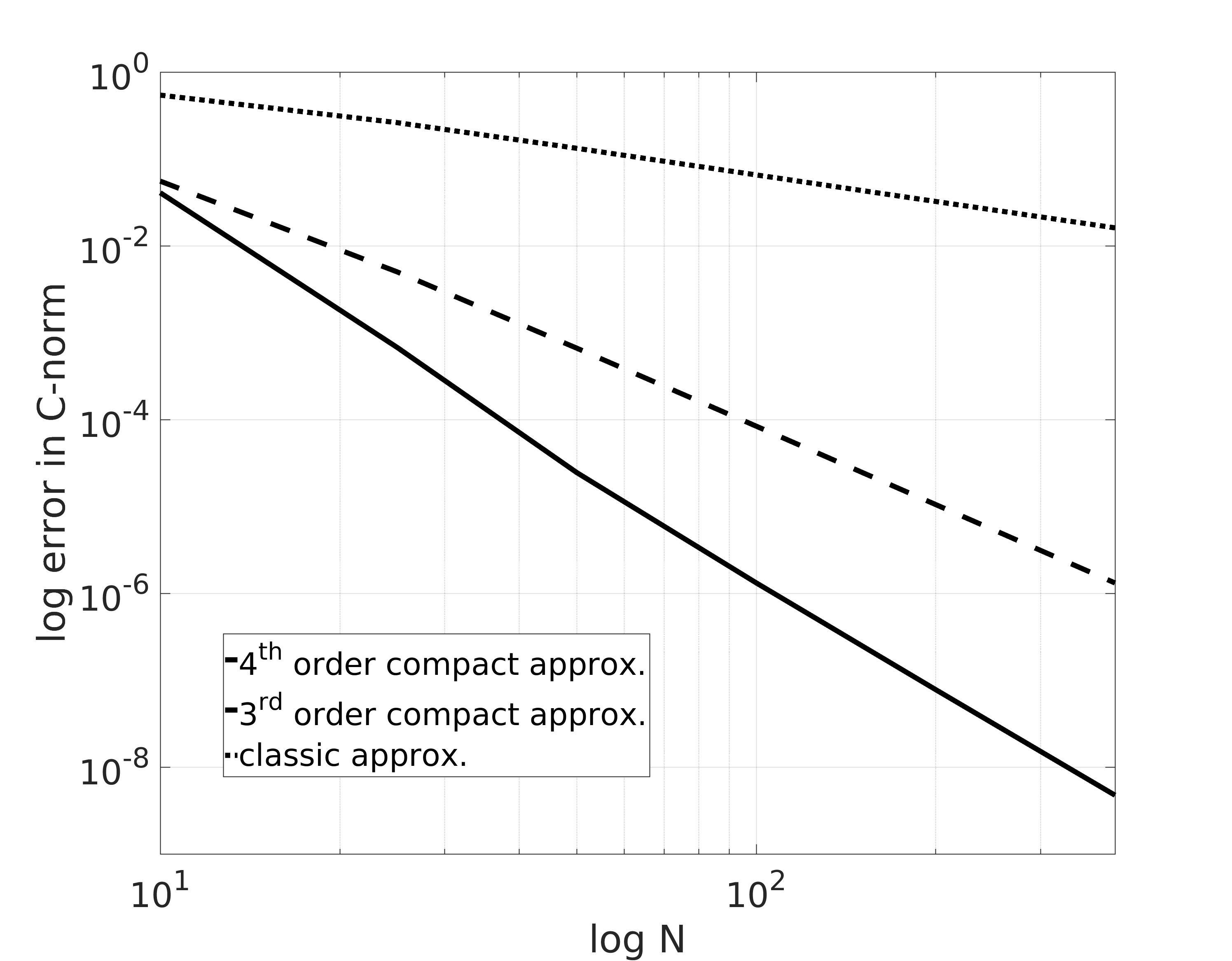}
    	\caption{
    Errors of the compact scheme on sample solutions (\ref{test_neumann}) of diffusion equation (\ref{diffusion_eq}) at $\nu^{\star} = 1,\: T = 1$. Bilogarithmic scale. Joint usage of compact difference scheme (\ref{ll_comp_scheme}) and boundary conditions approximation (\ref{neumann_coef}) shows the $4^{th}$ error decrease rate, while exploiting coefficients for the reduced two-point stencil results in $3^{rd}$ order. Classic approximation (with $\epsilon = 0.5$) decreases the error rate down to $1^{st}$.
    	}
	\label{fig:neumann3}
    \end{minipage}
\end{figure}

\begin{figure}
\begin{center}
	\includegraphics[scale=.5]{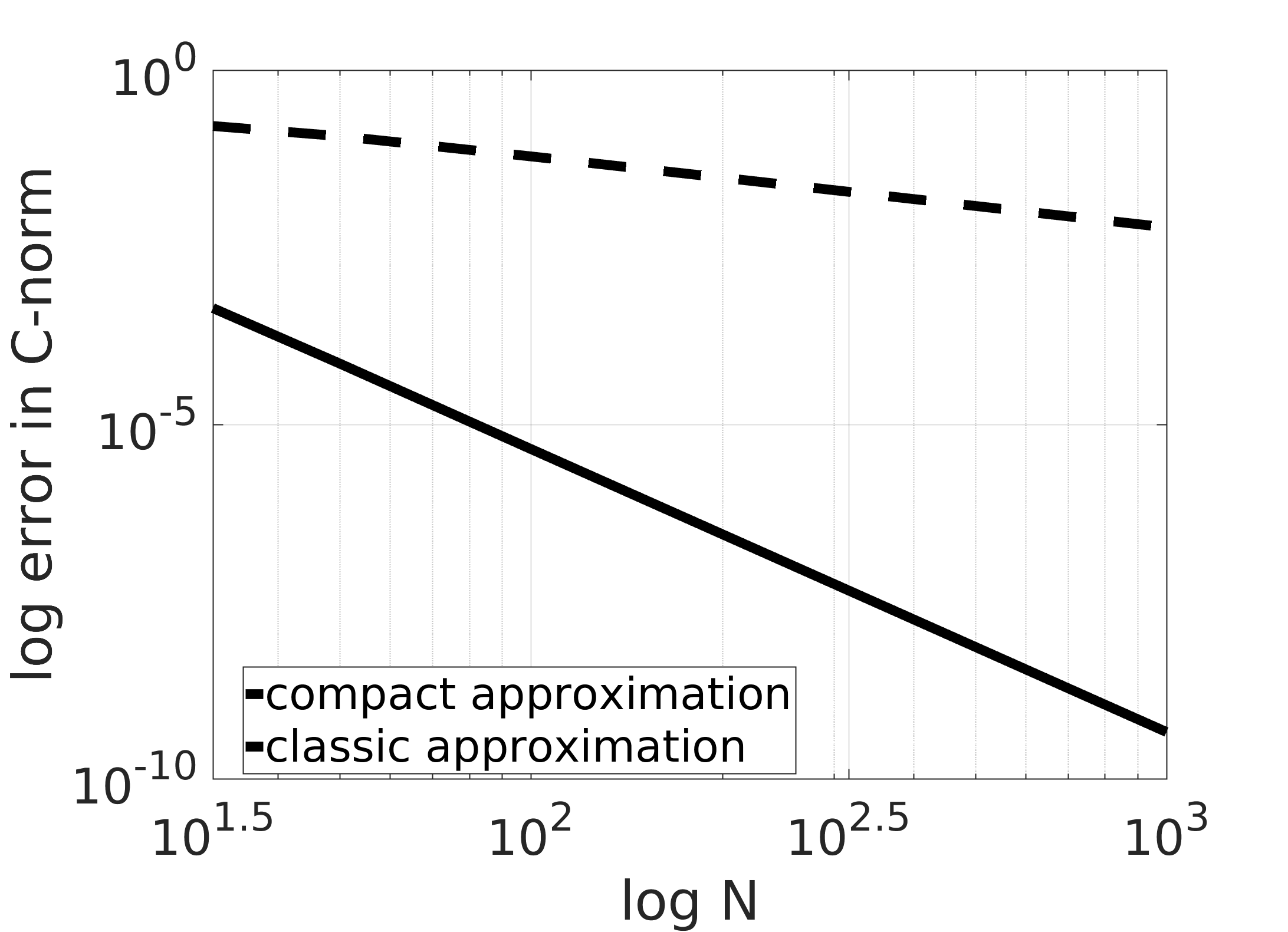}
	\caption{
Errors of compact scheme on sample solutions (\ref{test_neumann}) of Leontovich -- Levin equation (\ref{Schro_eq}) at $\nu^{\star} = 5i,\: T = 1$.  Bilogarithmic scale. Joint usage of compact difference scheme (\ref{ll_comp_scheme}) and boundary conditions approximation (\ref{neumann_coef}) show the $4^{th}$ error decrease rate, while classic approximation (with $\epsilon = 0.5$) decreases the error rate down to $1^{st}$. However, if one will use the main terms (limits of the coefficients as $h \rightarrow 0$) of compact approximation (\ref{neumann_coef}), thus cutting its coefficients, the error decrease rate will be equal to $3$.
	}
	\label{fig:neumann2}
\end{center}
\end{figure}

Numerical experiments confirm the $4-th$ order of approximation for the joint usage of the compact difference scheme (\ref{ll_comp_scheme}) and compact boundary conditions approximation (\ref{neumann_coef}), see Fig.~\ref{fig:neumann}, \ref{fig:neumann2}. We have constructed similar coefficients on the two-point stencil and reduced approximation order, i.e. with $\alpha^0_2 = \alpha^1_2 = \beta^0_2 = \beta^1_2 = 0$ and reduced basis of test functions $<u^{**}_{k_1, k_2}, f^{**}_{k_1, k_2}>$: $u^{**}_{k_1, k_2} \in \{ 1, t, t^2, x^2, x^2 t, x^2 t^2, x^3\}$; $f^{**}_{k_1, k_2} = \frac{\partial  u^{**}_{k_1, k_2}}{\partial t} - P  u^{**}_{k_1, k_2}$. Our numerical experiments show that the basis set reduction affects the approximation order, see Fig.~\ref{fig:neumann3}. This result differs from the one in \cite{britt2011numerical}, where the compact difference scheme was used to approximate the Helmholtz equation.

The classic approximation of Neumann boundary conditions is the following:
\begin{equation}\label{classic_neumann_bc}
    \epsilon (u^{n+1}_1 - u^{n+1}_0) + (1 - \epsilon) (u^{n}_1 - u^{n}_0) = 0, ~~0 < \epsilon \leq 1.
\end{equation}

Approximation (\ref{classic_neumann_bc}) provides the second order for a classic Crank -- Nicolson scheme (see Sect. \ref{sect:implicit-classic}), however, for the compact scheme (\ref{comp_scheme}) we obtain the first order only, see Fig. \ref{fig:neumann3}.

\subsection{Classic implicit scheme}\label{sect:implicit-classic}

We compare the compact scheme (\ref{comp_scheme}) and the results of our numerical experiments with classic (see e.g. \cite{samarskii-01}) 
ones.
The implicit second order finite difference scheme for (\ref{diffusion_eq}) can be
written in the following form:
\begin{equation} \label{divergent_1}
\frac{u^{n+1}_j - u^n_j}{\tau} = \frac{1}{2h^2}[u^n_{j+1}\theta_{j+\frac{1}{2}} + 
u^n_{j-1}\theta_{j-\frac{1}{2}} 
- u^n_j(\theta_{j+\frac{1}{2}} + \theta_{j-\frac{1}{2}}) + 
\end{equation}
\[
 + u^{n+1}_{j+1}\theta_{j+\frac{1}{2}} + u^{n+1}_{j-1}\theta_{j-\frac{1}{2}}
- u^{n+1}_j(\theta_{j+\frac{1}{2}} + \theta_{j-\frac{1}{2}})] + (f^{n+1}_j + f^n_j)/2 .
\]

The alternative versions for the right hand-side approximation are:
\[
(F^{n+1}_j + F^n_j)/2,
\]
where 
\begin{equation} \label{divergent_2}
F_j=(f_{j-1}+2f_j+f_{j+1})/4,
\end{equation}
or
\begin{equation} \label{divergent_3}
F_j=(f_{j-1}+2f_{j-1/2}+2f_j+2f_{j+1/2}+f_{j+1})/4.
\end{equation}

Our numerical experiments demonstrate very similar errors for these versions of the implicit scheme, see Table~\ref{tab:exper_accuracy}.

Implicit scheme (\ref{divergent_1}) is, in fact, a version of the well-known Crank -- Nicolson scheme. The scheme 
\begin{equation*}
\frac{u^{n+1} - u^n}{\tau} = A \frac{u^{n+1} + u^n}{2},
\end{equation*}
where $A$ is a negative definite self-adjoint operator is unconditionally stable in both finite- and infinite-dimensional cases.

Let us consider the transition operator for the implicit scheme in the case of $f(x)\equiv 0$. The 
matrices $A_{new}$ and $A_{old}$ (see Section \ref{matr_section}) are self-adjoint. They commute because their 
difference is proportional to the identity matrix. Therefore, like operators $R$ for (\ref{sturm}), the finite 
dimensional operator (matrix) $A_{new}^{-1}A_{old}$ is self-adjoint and contractive in the Euclidean 
metric $l^2$. 
Therefore, the unique limit $u^*_j=\lim\limits_{n\to +\infty} u^n_j$ exists  for any stationary forcing 
$f=f(x)$. The implicit scheme is unconditionally stable.

\subsection{The Leontovich -- Levin (Schr\"odinger-type) equation}
Compact scheme (\ref{comp_scheme}) can be modified to approximate the PDE
\begin{equation} \label{Schro_eq}
i \frac{\partial \Psi}{\partial t}=P\Psi+f,\quad P\Psi=
\frac{\partial}{\partial x}\theta(x) \frac{\partial \Psi}{\partial x},
\end{equation}
which is known as the Leontovich -- Levin equation, see e.g. a review \cite{leont-10}. Here $i$ is an imaginary unit, the solution 
$\Psi=\Psi(t,\,x)$ is a unknown complex-valued function, the positive function 
(coefficient) $\theta=\theta (x)$ is known, as well as the complex-valued function $f=f(x)$. This equation describes e.g. an electromagnetic 
field of the linear vibrators.

{\bf Note}. If the coefficient $\theta$ is constant, equation (\ref{Schro_eq}) is the famous 
Schr\"odinger  equation, see e.g. \cite{gord-10, gord-14}.

The operator $iP$ is skew self-adjoint in the space $L^2 [0,\,2\pi]$, and
therefore the resolving operator $\exp(iPt)$ of the mixed initial-boundary problem in the space is unitary.

In the case of $f \equiv 0$ the first integral of equation (\ref{Schro_eq}) under Dirichlet (or Neumann, or periodical)
boundary conditions exists:
\[
\int\limits_0^{2\pi} |\Psi (t,\,x)|^2\,dx=const.
\]

If we multiply the coefficients $\nu_j$ in Table~\ref{tab:coeffsu} by the imaginary unit $i$, we obtain the compact finite 
difference scheme, which approximates equation (\ref{Schro_eq}) with the $4$-th order; see its error in Table~\ref{tab:exper_accuracy_schrod}, where we compare its error with the error of the classic second order implicit scheme with the same temporal and spatial steps $\tau$ and $h$.

\section{Numerical experiments for parabolic equations}
\subsection{Sample solutions}
We use several sample solutions with various properties both for diffusion and for Leontovich -- Levin equations for various boundary conditions. These solutions were chosen to differ from the test functions used for scheme construction.

\begin{equation} \label{test_sol1}
\begin{split}
    u^*(t, x) & = sin^3(x)sin(t) + sin(2x)cos(t); \\
    \theta^*(x) & = cos^2(x) + 1. \\
\end{split}
\end{equation}

Hereafter we use the right-hand side
$f^*(t, x) = \frac{\partial u^*}{\partial t}-\frac{\partial}{\partial x}\theta^*(x) \frac{\partial u^*}{\partial x}$, i.e. our analytic sample solutions are exact.

\begin{figure}
    \centering
    \begin{minipage}{0.45\textwidth}
        \centering
        \includegraphics[scale=.45]{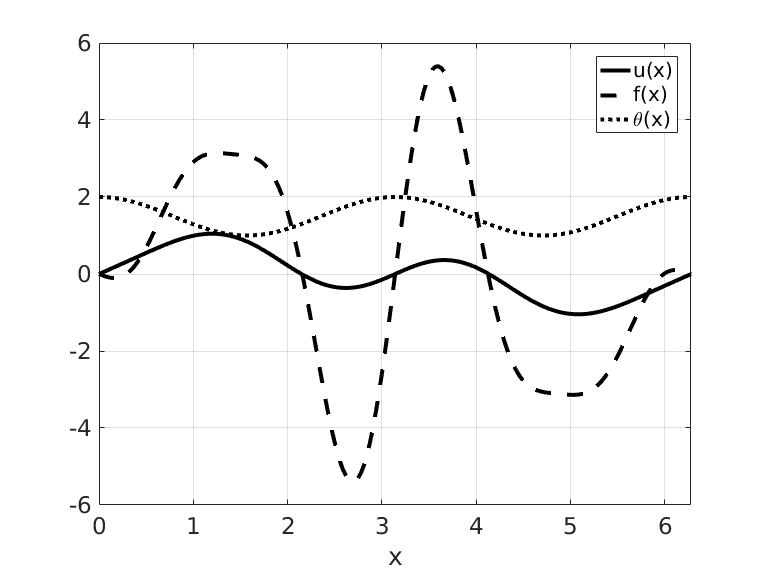} 
        \caption{
        Sample solution $u^*(T, x)$, right hand side $f^*(T, x)$ and coefficient $\theta^*(x)$ 
        (\ref{test_sol1}); $T = 1, N = 100$.
        	}\label{fig:sol1}
    \end{minipage}\hfill
    \begin{minipage}{0.45\textwidth}
        \centering
        \includegraphics[scale=.45]{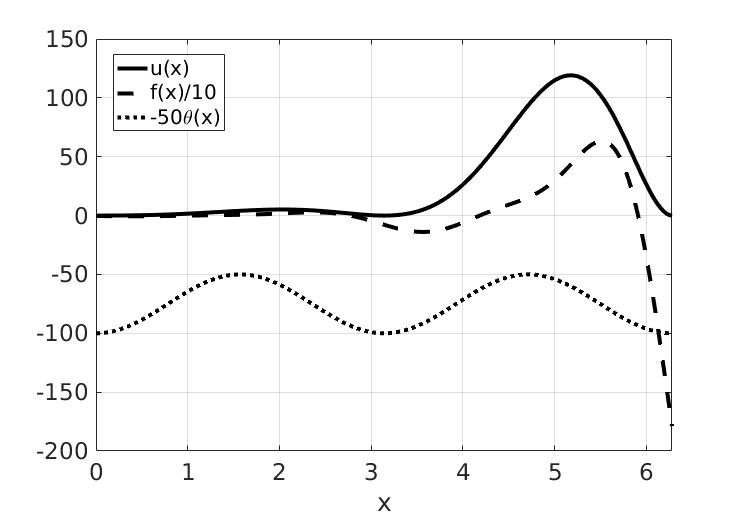} 
        \caption{
        Sample solution $u^*(T, x)$, right hand side $f^*(T, x)$ and coefficient $\theta^*(x)$ 
        (\ref{test_sol2}) at $k = 2$; $T = 1, N = 100$.
        	}\label{fig:sol2_k2}
    \end{minipage}
\end{figure}

\begin{figure}
    \centering
    \begin{minipage}{0.45\textwidth}
        \centering
        \includegraphics[scale=.45]{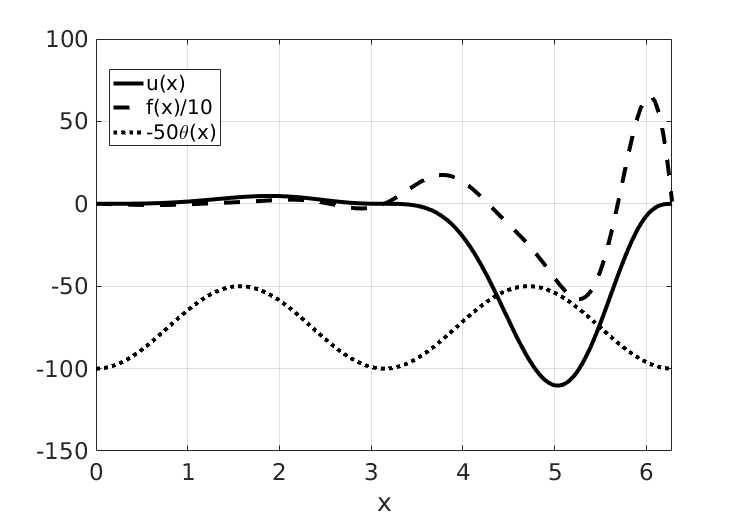} 
        \caption{
Sample solution $u^*(T, x)$, right hand side $f^*(T, x)$ and coefficient $\theta^*(x)$ (\ref{test_sol2}) at $k = 3$; $T = 1, N = 100$.
	}
    \label{fig:sol2_k3}
    \end{minipage}\hfill
    \begin{minipage}{0.45\textwidth}
        \centering
        \includegraphics[scale=.45]{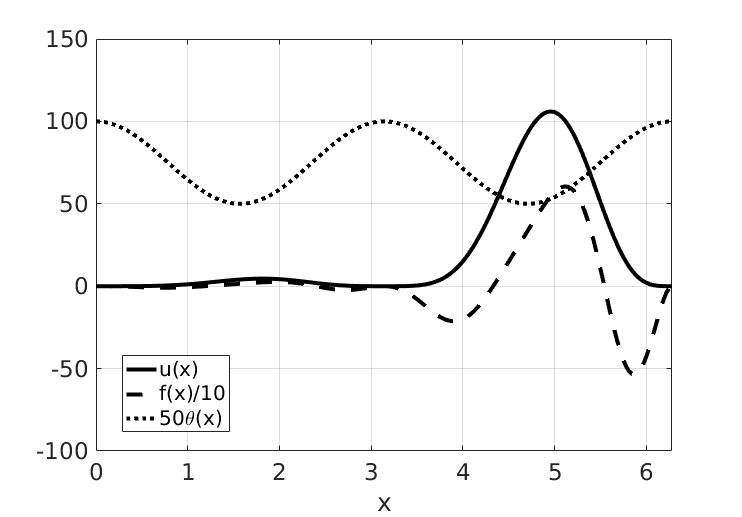} 
        \caption{
Sample solution $u^*(T, x)$, right hand side $f^*(T, x)$ and coefficient $\theta^*(x)$ 
(\ref{test_sol2}) at $k = 4$; $T = 1, N = 100$.
	}
	\label{fig:sol2_k4}
    \end{minipage}
\end{figure}

Then we consider the family of solutions which are very asymmetric with respect to $x$:

\begin{equation} \label{test_sol2}
\begin{split}
    u^*(t, x) & = sin(t)sin^k(x)exp(x); \\
    \theta^*(x) & = cos^2(x) + 1. \\
\end{split}
\end{equation}

Here the parameter $k$ in the sample solutions’ family controls their behavior near the endpoints $0$ and $2\pi$.





We also consider sample solutions with a very asymmetric coefficient $\theta(x)$:
\begin{equation} \label{test_sol3}
\begin{split}
    u^*(t, x) & = sin(x/2)(e^{b (2 \pi - x)}cos(\omega t) + e^{b x}sin(\omega t)); \\
    \theta^*(x) & = e^{a x}. 
\end{split}
\end{equation}


\begin{figure}[h!]
    \centering
    \begin{minipage}{0.45\textwidth}
        \centering
        \includegraphics[scale=.45]{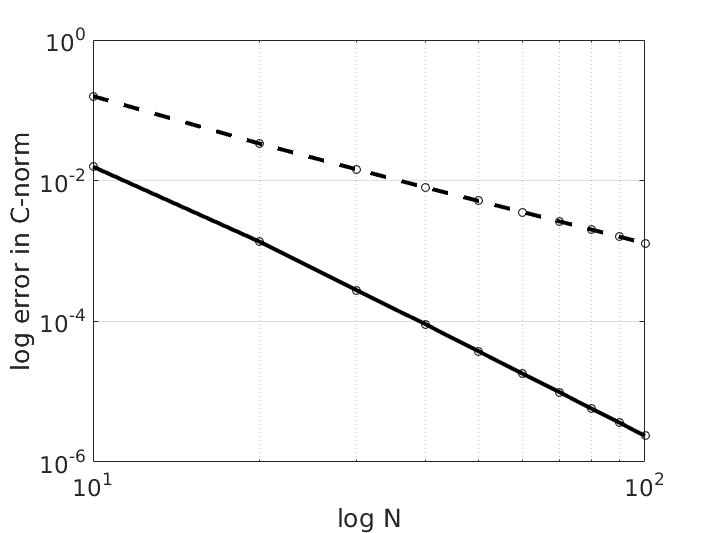} 
        \caption{
        Errors of compact and classic implicit schemes for sample solution (\ref{test_sol1}), 
        $\nu^{\star} = 1,\: T = 1$. Compact scheme (solid line) outperforms classic one (dashed line) both in 
        accuracy and convergence rate ($4-$th vs $2-$nd). Bilogarithmic scale.
    }
	\label{fig:error_smooth}
    \end{minipage}\hfill
    \begin{minipage}{0.45\textwidth}
        \centering
        \includegraphics[scale=.4]{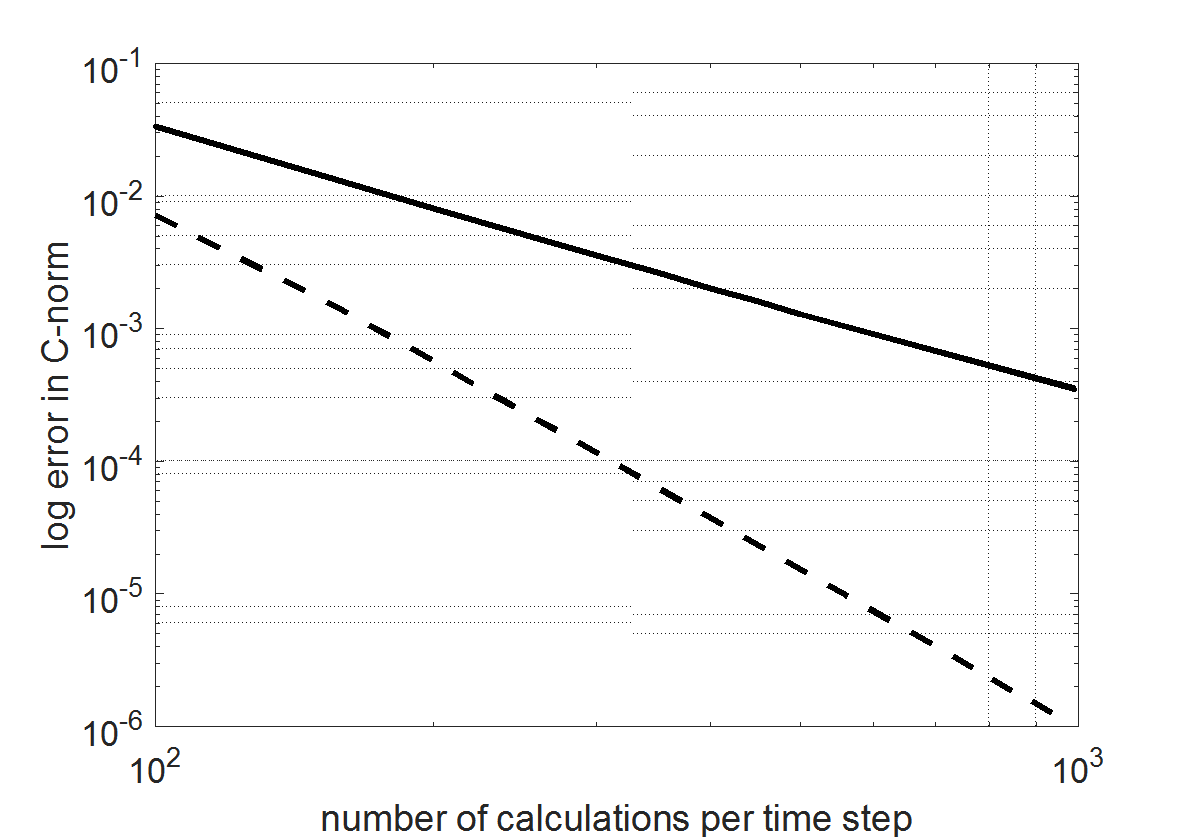} 
        \caption{
        Errors of compact and classic implicit schemes for sample solution (\ref{test_sol1}), 
        $\nu^{\star} = 1,\: T = 1$ as a function of number of operations (multiplications) per time step. Compact scheme (solid line) outperforms classic one (dashed line) both in 
        accuracy and convergence rate ($4-$th vs $2-$nd). Bilogarithmic scale.
    }
	\label{fig:error_eff}
    \end{minipage}
\end{figure}

We also consider the solution for the Neumann boundary problem for diffusion equation (\ref{diffusion_eq}):
\begin{equation} \label{test_neumann}
\begin{split}
    u^*(t, x) & = cos^2(x) sin (t); \\
    \theta^*(x) & = cos^2(x) + 1. 
\end{split}
\end{equation}

and for Leontovich -- Levin equation (\ref{Schro_eq}):
\begin{equation} \label{test_neumann_ll}
\begin{split}
    u^*(t, x) & = cos^2(x) sin (t); \\
    \theta^*(x) & = i(cos^2(x) + 1). 
\end{split}
\end{equation}

\subsection{Convergence rate of the scheme} 

\begin{figure}[h]
\begin{center}
	\includegraphics[scale=.5]{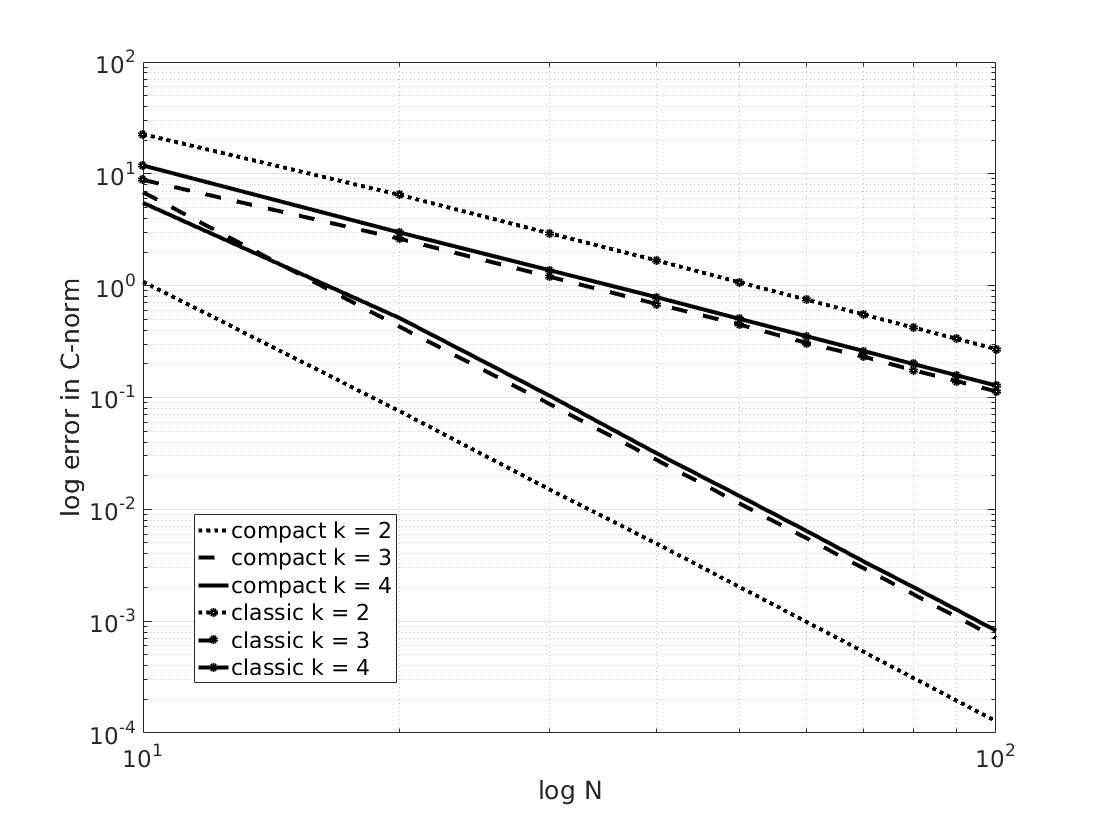}
	\caption{
Errors of the compact and classic implicit schemes on sample solutions (\ref{test_sol2}) of 
diffusion equation (\ref{diffusion_eq}) at $\nu^{\star} = 1,\: T = 1$. The compact scheme outperforms the classic 
one by both accuracy and convergence rate for sample solutions 
(\ref{test_sol2}) with various values of $k$. Bilogarithmic scale.
	}
	\label{fig:error_ks}
\end{center}
\end{figure}

We find approximate solutions by the compact scheme (\ref{comp_scheme}) and by the classic implicit 
scheme (\ref{divergent_1}) on sample solutions (\ref{test_sol1}-\ref{test_neumann_ll}), see 
Fig.~\ref{fig:sol1}-\ref{fig:sol2_k4}. To measure the convergence rate, we fix a Courant parameter $\nu^{\star}$, thus determining $\tau =h^2|\nu^{\star}|/\max\limits_{j} \theta_j$ for every $N$. We can see that compact scheme (\ref{comp_scheme}) gives us a
much smaller error than the classic implicit one, see Fig.~\ref{fig:error_smooth}, \ref{fig:error_ks} and 
Table~\ref{tab:exper_accuracy}. The error of the compact scheme in these solutions is much smaller; 
the compact scheme demonstrates an error of the $4$-th order vs. the second order for the classic scheme.

\begin{table}
\centering
\caption{Error in {\bf C}-norm for various sample solutions of equation~\ref{diffusion_eq}). Here $\nu^{\star} = 1,\: T = 1$. Compact scheme outperforms all the 
variations of implicit scheme by both accuracy and order. Among implicit schemes 
right-hand side averaging variants, namely (\ref{divergent_1}) shows the best accuracy.}
\label{tab:exper_accuracy}
\begin{tabular}{|c|c|c|c|c|c|c|}
\hline
Sample solution & Scheme & N = 10 & N = 20 & N = 50 & N = 100 & order \\  \hline
(\ref{test_sol1}) & (\ref{comp_scheme}) & 1.58-2 & 1.36-3 & 3.73-5 & 2.36-6 & 3.83 \\  \hline
(\ref{test_sol1}) & (\ref{divergent_1}) & 1.59-1 & 3.38-2 & 5.14-3 & 1.29-3 & 2.09 \\  \hline
(\ref{test_sol1}) & (\ref{divergent_2}) & 2.18-1 & 4.94-2 & 8.67-3 & 2.16-3 & 2.00 \\  \hline
(\ref{test_sol1}) & (\ref{divergent_3}) & 2.18-1 & 7.38-2 & 1.40-2 & 3.67-3 & 1.93 \\  \hline
(\ref{test_sol2}), k = 2  & (\ref{comp_scheme}) & 1.09+0 & 7.53-2 & 2.03-3 & 1.28-4 & 3.93 \\  \hline
(\ref{test_sol2}), k = 2  & (\ref{divergent_1}) & 2.26+1 & 6.48+0 & 1.07+0 & 2.71-1 & 1.92 \\  \hline
(\ref{test_sol2}), k = 2  & (\ref{divergent_2}) & 4.74+1 & 1.47+1 & 2.49+0 & 6.28-1 & 1.99 \\  \hline
(\ref{test_sol2}), k = 2  & (\ref{divergent_3}) & 3.17+1 & 1.10+0 & 2.03+0 & 5.20-1 & 1.96 \\  \hline
(\ref{test_sol2}), k = 3  & (\ref{comp_scheme}) & 6.84+0 & 4.28-1 & 1.13-2 & 7.12-4 & 3.98 \\  \hline
(\ref{test_sol2}), k = 3  & (\ref{divergent_1}) & 8.83+0 & 2.63+0 & 4.50-1 & 1.13-1 & 1.89 \\  \hline
(\ref{test_sol2}), k = 3  & (\ref{divergent_2}) & 1.68+1 & 6.28+0 & 1.12+0 & 2.83-1 & 1.98 \\  \hline
(\ref{test_sol2}), k = 3  & (\ref{divergent_3}) & 1.06+1 & 3.74+0 & 7.76-1 & 2.11-1 & 1.88 \\  \hline
(\ref{test_sol2}), k = 4  & (\ref{comp_scheme}) & 5.50+0 & 5.12-1 & 1.32-2 & 8.22-4 & 3.83 \\  \hline
(\ref{test_sol2}), k = 4  & (\ref{divergent_1}) & 1.19+1 & 2.99+0 & 5.06-1 & 1.28-1 & 1.97 \\  \hline
(\ref{test_sol2}), k = 4  & (\ref{divergent_2}) & 2.68+1 & 7.27+0 & 1.27+0 & 3.19-1 & 1.99 \\  \hline
(\ref{test_sol2}), k = 4  & (\ref{divergent_3}) & 1.16+1 & 3.47+0 & 7.42-1 & 1.94-1 & 1.94 \\  \hline
\end{tabular}
\end{table}

\begin{table}
\centering
\caption{Error in {\bf C}-norm for sample solution (\ref{test_sol3}) with various parameters. Here $\nu^{\star} = 100,\: T = 1$. Compact scheme outperforms classic implicit scheme (\ref{divergent_1}) by both accuracy and order.}
\label{tab:exper_accuracy_3}
\begin{tabular}{|c|c|c|c|c|c|c|}
\hline
Parameters & Scheme & N = 10 & N = 20 & N = 50 & N = 100 & order \\  \hline
$a = 1, b = 1, \omega = 1$ & (\ref{divergent_1}) & 1.99+1 & 5.69+0 & 9.38-1 & 2.35-1 & 1.95 \\  \hline
$a = 1, b = 1, \omega = 1$ & (\ref{comp_scheme}) & 6.59-1 & 4.60-2 & 1.20-3 & 7.55-5 & 3.96 \\  \hline
$a = 1, b = 2, \omega = 2$ & (\ref{divergent_1}) & 1.93+4 & 5.45+3 & 9.02+2 & 2.27+2 & 1.95 \\  \hline
$a = 1, b = 2, \omega = 2$ & (\ref{comp_scheme}) & 3.73+3 & 2.47+2 & 6.41+0 & 4.02-1 & 3.98 \\  \hline
$a = 2, b = 1, \omega = 1$ & (\ref{divergent_1}) & 3.02+1 & 8.56+0 & 1.42+0 & 3.57-1 & 1.95 \\  \hline
$a = 2, b = 1, \omega = 1$ & (\ref{comp_scheme}) & 9.74-1 & 6.18-2 & 1.59-3 & 9.92-5 & 3.99 \\  \hline
$a = 1, b = 0.1, \omega = 1$ & (\ref{divergent_1}) & 5.22-2 & 1.38-2 & 2.22-3 & 5.56-4 & 1.98 \\  \hline
$a = 1, b = 0.1, \omega = 1$ & (\ref{comp_scheme}) & 7.47-5 & 3.99-6 & 9.90-8 & 6.17-9 & 4.06 \\  \hline
$a = 1, b = 2, \omega = 10$ & (\ref{divergent_1}) & 1.25+4 & 1.86+3 & 4.25+2 & 1.12+2 & 2.02 \\  \hline
$a = 1, b = 2, \omega = 10$ & (\ref{comp_scheme}) & 2.60+3 & 1.10+2 & 2.71+0 & 1.78-1 & 4.09 \\ \hline
\end{tabular}
\end{table}



\subsection{Efficiency} 

In 1D case, the double-sweep method for tridiagonal matrix requires $5N$ multiplications and divisions. On every time step, classic scheme (\ref{divergent_1}) requires double-sweep only, while usage of a compact scheme involves additional $3N$ multiplications for the right-hand side. We thus compare both schemes in terms of efficiency in a numerical experiment, see Fig.~\ref{fig:error_eff}. Compact difference scheme outperforms the classic one in experimental setup even on efficiency-based comparison with number of calculations per time step fixed.

In the multidimensional case, iterative methods are widely used and are most effective. The matrix for the left hand-side is block-tridiagonal for both compact and classic schemes, the difference is the modified right-hand side, which is multiplied by another block-tridiagonal matrix. Since inversion of the left-hand side matrix is much more time-consuming, we hope that our approach will be also effective for the multidimensional case; this, however, require further investigation and numerical experiments.

\subsection{Richardson extrapolation}
The extrapolation Richardson method may be used to improve the schemes order. If we obtain a family
 of approximate solutions $u=u_{h}(t,\,x)$ at $t=T$, assume $\tau =h^2|\nu^{\star}|/\max\limits_{j} \theta_j$, 
and use the representation
\begin{equation}\label{Ri}
u_h (T,\,x)= u(T,\,x)+h^4 u_*(T,\,x) ,   
\end{equation}
then we can calculate $u_h$ at $h=h_*$ and at $h=h_*/2$. Afterwards, we substitute the representation into (\ref{Ri}), and obtain a simple linear algebraic system:
\begin{equation*}
u_{h_\star}=u+h^4_\star u_\star,\; u_{h_*/2}=u+h^4_\star u_\star/16 \rightarrow u=u(T,x)\approx \left[ 16 u_{h_\star/2}- u_{h_\star}\right]/15.
\end{equation*}

\begin{table}[h]
\centering
\caption{Error in {\bf C}-norm for various sample solutions using Richardson extrapolation for scheme (\ref{comp_scheme}). 
Compact scheme outperforms classic implicit scheme in both accuracy and convergence rate.}
\label{tab:exper_accuracy_richardson}
\begin{tabular}{|c|c|c|c|c|c|c|}
\hline
Test solution & Scheme & N = 10 & N = 20 & N = 50 & N = 100 & order \\  \hline
(\ref{test_sol1}) & (\ref{divergent_1}) & 5.76-3 & 3.13-4 & 8.60-6 & 5.36-7 & 4.00 \\  \hline
(\ref{test_sol1}) & (\ref{comp_scheme}) & 1.31-4 & 2.35-6 & 9.30-9 & 1.44-10 & 6.01 \\  \hline
(\ref{test_sol2}), k = 2 & (\ref{divergent_1}) & 4.77-1 & 3.72-2 & 9.91-4 & 6.29-5 & 3.98 \\  \hline
(\ref{test_sol2}), k = 2 & (\ref{comp_scheme}) & 8.10-3 & 2.26-4 & 9.27-7 & 1.46-8 & 5.99 \\  \hline
(\ref{test_sol2}), k = 3 & (\ref{divergent_1}) & 2.10+0 & 9.40-2 & 2.47-3 & 1.54-4 & 4.00 \\  \hline
(\ref{test_sol2}), k = 3 & (\ref{comp_scheme}) & 1.34-1 & 1.60-3 & 6.15-6 & 9.55-8 & 6.01 \\  \hline
(\ref{test_sol2}), k = 4 & (\ref{divergent_1}) & 1.94+0 & 1.52-1 & 3.80-3 & 2.37-4 & 4.00 \\  \hline
(\ref{test_sol2}), k = 4 & (\ref{comp_scheme}) & 3.74-2 & 2.68-3 & 1.02-5 & 1.59-7 & 6.00 \\  \hline
\end{tabular}
\end{table}

\begin{figure}[h!]
\begin{center}
	\includegraphics[scale=.7]{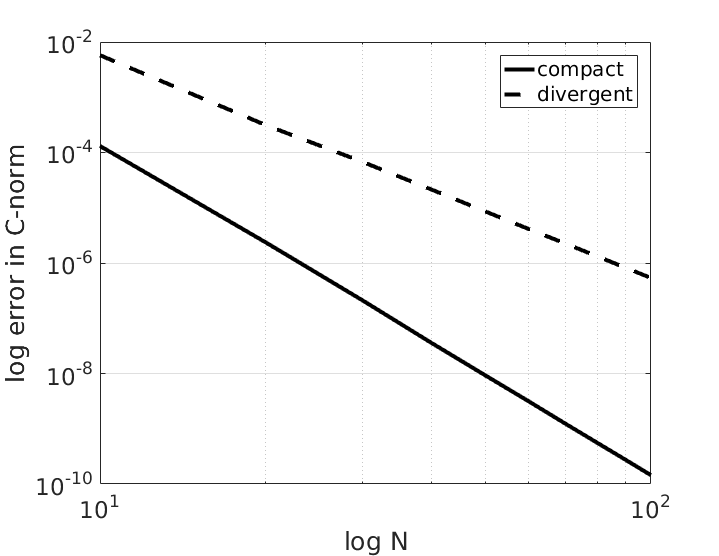}
	\caption{
Errors of compact and classic implicit schemes in sample solution (\ref{test_sol1}). 
Richardson extrapolation technique is used here to improve the convergence rate.
Compact scheme (solid line) outperforms the classic one (dashed line) both by accuracy and convergence rate ($4-$th vs $6-$th). Bilogarithmic scale.
	}
	\label{fig:richardson}
\end{center}
\end{figure}

We can improve (by using representation (\ref{Ri})) the order of both schemes: we improve the order 
for classic implicit scheme~(\ref{cscheme}) up to $4$-th and for compact scheme up~(\ref{comp_scheme})to $6$-th, see the 
Fig.~\ref{fig:richardson} and Table~\ref{tab:exper_accuracy_richardson}.

\subsection{Cut coefficients}

If we eliminate in the coefficients of compact scheme (\ref{comp_scheme}) terms with a power 
of $h$ more than $4$ (e.g. $h^5$, $h^6$, \ldots , see Table \ref{tab:coeffsu}, \ref{tab:coeffsf}), the $4$-th order will be preserved. However, the 
absolute error may increase a bit, see Table \ref{tab:cut_accuracy}.

\begin{table}[h!]
\centering
\caption{Error in {\bf C}-norm for various cases of coefficients of the compact scheme (\ref{diffusion_eq}) with terms cut for sample solution (\ref{test_sol1}). $\nu^{\star} =1,\: T = 1$.}
\label{tab:cut_accuracy}
\begin{tabular}{|c|c|c|c|c|c|}
\hline
Cutting terms with & N = 10 & N = 20 & N = 50 & N = 100 & order \\  \hline
$h^5$ and greater & 1.5659-2 & 1.7958-3 & 5.5527-5 & 3.5936-6 & 3.95 \\  \hline
$h^6$ and greater & 1.6422-2 & 1.1773-3 & 3.6297-5 & 2.3422-6 & 3.95 \\  \hline
$h^7$ and greater & 1.5917-2 & 1.3402-3 & 3.7068-5 & 2.3563-6 & 3.98 \\  \hline
$h^8$ and greater & 1.6026-2 & 1.3750-3 & 3.7280-5 & 2.3594-6 & 3.98 \\  \hline
$h^9$ and greater & 1.5965-2 & 1.3651-3 & 3.7272-5 & 2.3593-6 & 3.98 \\  \hline
Exact scheme & 1.5812-2 & 1.3642-3 & 3.7271-5 & 2.3593-6 & 3.98 \\  \hline
\end{tabular}
\end{table}

\subsection{Almost self-adjoint matrices} \label{matr_section}

Our finite difference scheme may be rewritten in the matrix (operator) form:
\begin{equation*}
A_{new}u^{n+1} + A_{old}u^n = B_{old}f^n + B_{new}f^{n+1},   
\end{equation*}
 where

$A_{new} = 
 \begin{pmatrix}
  1 & 0 & 0 & 0 & \cdots & 0 & 0 \\
  b^L_{1, 2} & a_{1, 2} & b^R_{1, 2} & 0 & \cdots & 0 & 0 \\
  0 & b^L_{1, 3} & a_{1, 3} & b^R_{1, 3} & \cdots & 0 & 0 \\
  \vdots & \vdots & \vdots & \vdots & \ddots & \vdots & \vdots  \\
  0 & 0 & 0 & 0 & \cdots & 0 & 0 \\
  0 & 0 & 0 & 0 & \cdots & a_{1, N-1} & b^R_{1, N-1} \\
  0 & 0 & 0 & 0 & \cdots & 0 & 1 
 \end{pmatrix}$,
\\
$A_{old} = 
 \begin{pmatrix}
  0 & 0 & 0 & 0 & \cdots & 0 & 0 \\
  b^L_{0, 2} & a_{0, 2} & b^R_{0, 2} & 0 & \cdots & 0 & 0 \\
  0 & b^L_{0, 3} & a_{0, 3} & b^R_{0, 3} & \cdots & 0 & 0 \\
  \vdots & \vdots & \vdots & \vdots & \ddots & \vdots & \vdots  \\
  0 & 0 & 0 & 0 & \cdots & 0 & 0 \\
  0 & 0 & 0 & 0 & \cdots & a_{0, N-1} & b^R_{0, N-1} \\
  0 & 0 & 0 & 0 & \cdots & 0 & 0
 \end{pmatrix}$,
\\

$B_{new} = 
 \begin{pmatrix}
  0 & 0 & 0 & 0 & \cdots & 0 & 0 \\
  q^L_{1, 2} & p_{1, 2} & q^R_{1, 2} & 0 & \cdots & 0 & 0 \\
  0 & q^L_{1, 3} & p_{1, 3} & q^R_{1, 3} & \cdots & 0 & 0 \\
  \vdots & \vdots & \vdots & \vdots & \ddots & \vdots & \vdots  \\
  0 & 0 & 0 & 0 & \cdots & 0 & 0 \\
  0 & 0 & 0 & 0 & \cdots & p_{1, N-1} & q^R_{1, N-1} \\
  0 & 0 & 0 & 0 & \cdots & 0 & 0 
 \end{pmatrix}$,
\\
$B_{old} = 
 \begin{pmatrix}
  0 & 0 & 0 & 0 & \cdots & 0 & 0 \\
  q^L_{0, 2} & p_{0, 2} & q^R_{0, 2} & 0 & \cdots & 0 & 0 \\
  0 & q^L_{0, 3} & p_{0, 3} & q^R_{0, 3} & \cdots & 0 & 0 \\
  \vdots & \vdots & \vdots & \vdots & \ddots & \vdots & \vdots  \\
  0 & 0 & 0 & 0 & \cdots & 0 & 0 \\
  0 & 0 & 0 & 0 & \cdots & p_{0, N-1} & q^R_{0, N-1} \\
  0 & 0 & 0 & 0 & \cdots & 0 & 0 
 \end{pmatrix}$.

We cannot prove our hypothesis: compact scheme (\ref{comp_scheme}) is unconditionally stable and convergent for any smooth and positive coefficient $\theta(x)$. However, we checked in multiple numerical experiments the following properties of the scheme for various cases.
 
The matrix $M=-A^{-1}_{new}A_{old}$ is not self-adjoint. However, all the eigenvalues $\left\{\lambda_j\right\}_{j=1}^N$ of matrix $M$ are real. The stability
condition $|\lambda_j|<1$ holds, and $M$ does not have non-trivial Jordan blocks for any values $\nu$ and for various boundary conditions. Therefore, there is an Euclidean norm, where the linear operator with the matrix $M$ is self-adjoint. 

We evaluate the distance between the matrix $M$ and the subspace of self-adjoint (symmetrical) matrices. Let us define for an arbitrary square matrix $C \in \mathbb{R}^{(N-1) \times (N-1)}$ its measure of asymmetry: 
\begin{equation} \label{asymm_measure}
    S(C) = \frac{|| C - C^*||_F}{N-1}.    
\end{equation}

\begin{figure}[h!]
    \centering
    \begin{minipage}{0.35\textwidth}
       \centering
        \includegraphics[scale=.4]{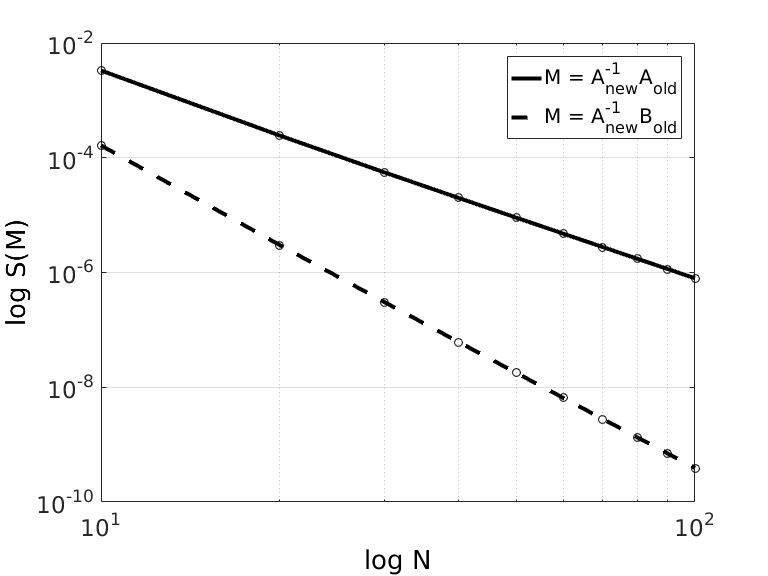} 
        \caption{
$S(A^{-1}_{new}A_{old})$ (solid line) and $S(A^{-1}_{new}B_{old})$ (dashed line) as a function of 
$N$ on sample solution (\ref{test_sol2}). Bilogarithmic scale.
	}
	\label{fig:symm_error}
    \end{minipage}
    \hfill
    \centering
    \begin{minipage}{0.55\textwidth}
        \includegraphics[scale=.55]{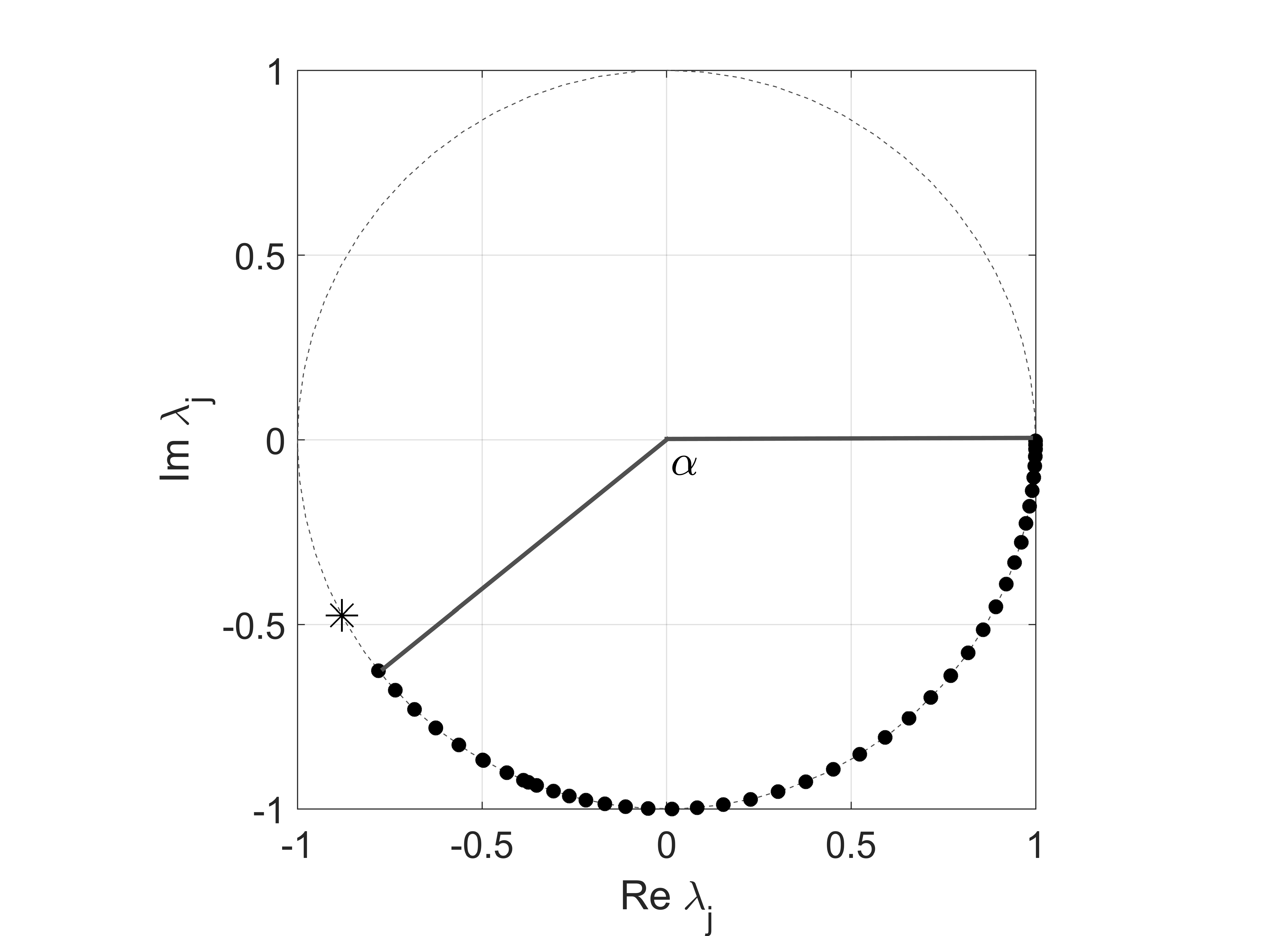} 
        \caption{
Eigenvalues $\lambda_j$ of the transition operator $M=-A^{-1}_{new}A_{old}$ for Leontovich -– Levin equation (\ref{Schro_eq}) on the complex plane for the Dirichlet boundary conditions. 
Dashed line shows the unit circle, i.e. all these eigenvalues are unimodal. 
If $\nu$ is fixed and $N \to \infty$, then the angle $\alpha \to const$. If $\nu \to 0$, then $ \alpha \to 0$; 
if $\nu \to \infty$, then $\alpha \to \pi$. In case of compact Neumann boundary conditions approximation (\ref{neumann_coef}), two coinciding eigenvalues appear, depicted by a star. Here $N = 50$, $\theta(x) = i [cos^2(x)+1]$, $\nu^{\star} = 1$.
	}
	\label{fig:eigens}
    \end{minipage}
\end{figure}

Here $||\cdot||_F$ is a Frobenius norm in the $N^2$-dimensional space of matrices. The measure $S(M)$ decreases
as $N\to\infty$, see Table~\ref{tab:exper_symm} and Fig.~\ref{fig:symm_error}. For the classic Euclidean norm $l^2$, the operator is asymptotically almost self-adjoint: $S(M)\approx N^{-4}$ as $N \rightarrow \infty$, see Table~\ref{tab:exper_symm}. 

\begin{figure}[h!]
\includegraphics[scale=.45]{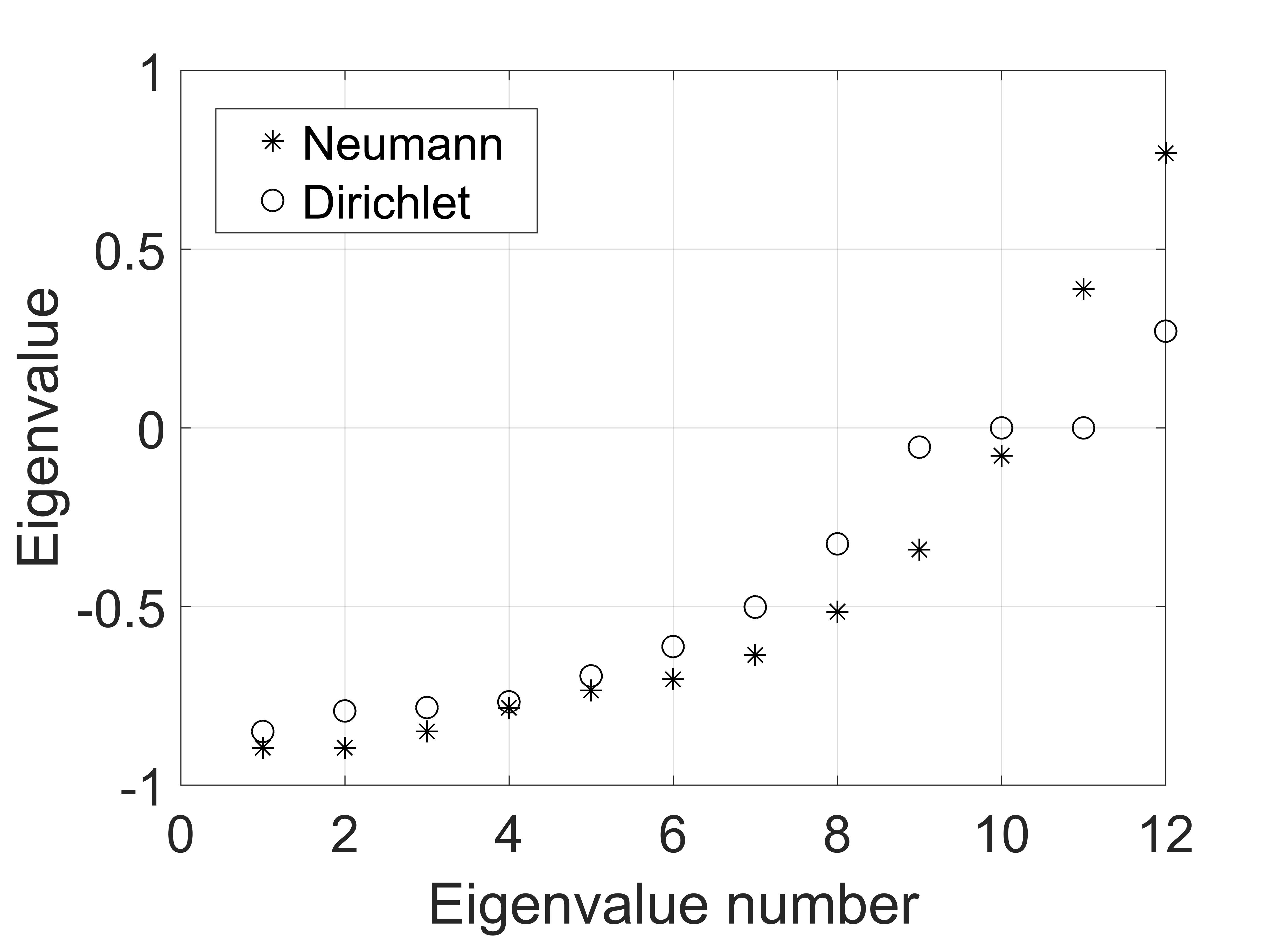} 
        \caption{
Eigenvalues $\lambda_j$ of the transition operator $M=-A^{-1}_{new}A_{old}$ for the diffusion equation (\ref{diffusion_eq}) for the Dirichlet and Neumann boundary conditions. Choice of compact Neumann boundary conditions approximation (\ref{neumann_coef}) extends the spectre.
Here $N = 12$, $\theta(x) = cos^2(x)+1$, $\nu^{\star} = 5$.
	}
	\label{fig:eigens_diff}
\end{figure}

\textbf{Note.} The spectrum $Spec~R(t)$ at $t > 0$ of resolving operators for differential problem (\ref{sturm}) is strongly negative under the Dirichlet boundary conditions. On the subspace of grid functions $u$ such that $u_0 = u_N = 0$, the spectrum of the finite-difference transition operator $M_{Dirichlet}$ is strongly negative at $\nu < \nu_\#$ only, where $\nu_\#  \approx 1/3$. This estimate is a result of our numerical experiments, see Fig. \ref{fig:eigens_diff}. In the case of the Neumann conditions $Spec~R(t)$ is non-positive. The non-positiveness for the operator $M_{Neumann}$ is fulfilled at $\nu_\# \approx 1/4$ only. In all the cases the $Spec~M_{Neumann}$ is wider than $Spec~M_{Dirichlet}$.

\begin{table}[h!]
\centering
\caption{Measure of asymmetry (\ref{asymm_measure}) for finite difference operators of the compact scheme (\ref{comp_scheme})}
\label{tab:exper_symm}
\begin{tabular}{|c|c|c|c|c|c|}
\hline
- &   $N = 10$ & $N = 20$ & $N = 50$ & $N = 100$ & order 		\\  \hline
$S(A^{-1}_{new}A_{old})$ & 3.32-3 & 2.44-4 & 9.05-6 & 7.93-7 & 3.62  \\  \hline
$S(A^{-1}_{new}B_{old})$ & 1.64-4 & 3.01-6 & 1.79-8 & 3.91-10 & 5.62  \\  \hline
\end{tabular}
\end{table}

\subsection{Numerical experiments for the Leontovich -- Levin \\ (Schr\"odinger--type) equation}

For the Leontovich -- Levin equation (\ref{Schro_eq}) we use the following scheme on six-point two-layer stencils (see 
Fig.~\ref{fig:stencil}):
\begin{equation}  \label{ll_comp_scheme}
\begin{aligned}
&B^L_{0,j}u^n_{j-1} + A_{0,j}u^n_j + B^R_{0,j}u^n_{j+1} + B^L_{1,j}u^{n+1}_{j-1} + A_{1,j}u^{n+1}_j + B^R_{1,j}u^{n+1}_{j+1} = 
\\
&B^L_{0,j}f^n_{j-1} + P_{0,j}f^n_j + Q^R_{0,j}f^n_{j+1} + Q^L_{1,j}f^{n+1}_{j-1} + P_{1,j}f^{n+1}_j + Q^R_{1,j}f^{n+1}_{j+1},    
\end{aligned}
\end{equation}
$j = 1, \ldots, N-1, \; n = 0, \ldots, T/\tau$. 
Here coefficients $A_{0,j}, A_{1,j}$, $B^L_{0,j}, B^L_{1,j}$, $B^R_{0,j}, B^R_{1,j}$, $P_{0,j}, 
P_{1,j}$, $Q^L_{0,j}, Q^L_{1,j}$, $Q^R_{0,j}, Q^R_{1,j}$ are the same as lowercase coefficients in 
Table ~\ref{tab:coeffsu}, ~\ref{tab:coeffsf}, but all the $\nu_j$ entrances should be multiplied by an imaginary unit $i$.

\begin{table}[h]
\centering
\caption{Error in {\bf C}-norm for various sample solutions for Leontovich -- Levin equation 
(\ref{Schro_eq}). Here $\nu^{\star} = i,\: T = 1$. The compact scheme outperforms implicit scheme 
(\ref{divergent_1}) by both accuracy and order.}
\label{tab:exper_accuracy_schrod}
\begin{tabular}{|c|c|c|c|c|c|c|}
\hline
Test solution & Scheme & N = 10 & N = 20 & N = 50 & N = 100 & order \\  \hline
(\ref{test_sol1}) & (\ref{divergent_1}) & 2.18-1 & 4.28-2 & 5.79-3 & 1.29-3 & 2.17 \\  \hline
(\ref{test_sol1}) & (\ref{comp_scheme}) & 2.58-2 & 1.86-3 & 5.12-5 & 3.22-6 & 3.99 \\  \hline
(\ref{test_sol2}), k = 2 & (\ref{divergent_1}) & 2.84+1 & 8.00+0 & 1.34+0 & 3.37-1 & 1.99 \\  \hline
(\ref{test_sol2}), k = 2 & (\ref{comp_scheme}) & 1.05+0 & 7.51-2 & 2.07-3 & 1.30-4 & 3.99 \\  \hline
(\ref{test_sol2}), k = 3 & (\ref{divergent_1}) & 1.27+1 & 3.25+0 & 5.23-1 & 1.33-1 & 1.98 \\  \hline
(\ref{test_sol2}), k = 3 & (\ref{comp_scheme}) & 8.50+0 & 5.67-1 & 1.47-2 & 9.23-4 & 4.00 \\  \hline
(\ref{test_sol2}), k = 4 & (\ref{divergent_1}) & 1.28+1 & 3.54+0 & 6.09-1 & 1.55-1 & 1.97 \\  \hline
(\ref{test_sol2}), k = 4 & (\ref{comp_scheme}) & 5.69+0 & 5.33-1 & 1.40-2 & 8.71-4 & 4.00 \\  \hline
\end{tabular}
\end{table}

\begin{figure}
\begin{center}
	\includegraphics[scale=.6]{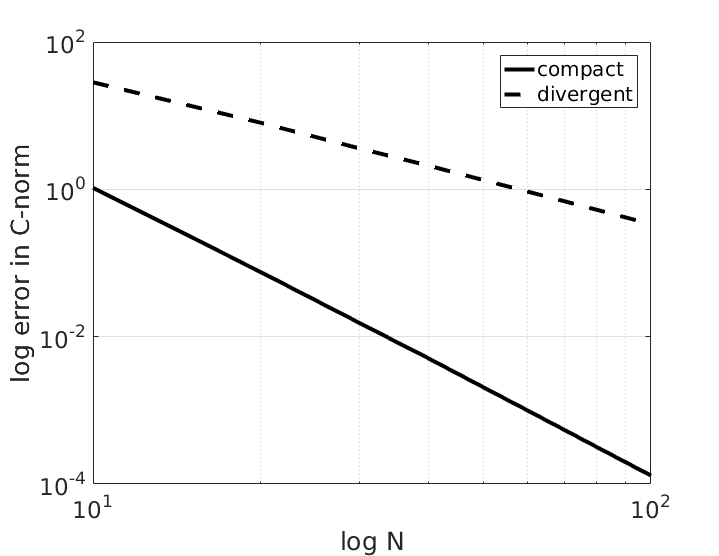}
	\caption{
Errors of compact and classic implicit schemes in sample solution (\ref{test_sol1}) of Leontovich -– Levin equation (\ref{Schro_eq}), 
$\nu^{\star} =i,\: T = 1$. Compact scheme (solid line) outperforms the classic one (dashed line) both in accuracy and order ($2$-nd vs $4$-th). Bilogarithmic scale.
	}
	\label{fig:schrod}
\end{center}
\end{figure}

\begin{table}[h]
\centering
\caption{Error in {\bf C}-norm for sample solution (\ref{test_sol3}) of Leontovich -- Levin equation (\ref{Schro_eq}). $\nu^{\star} = 100i,\: T = 1$. Compact scheme outperforms the implicit scheme (\ref{divergent_1}) by both accuracy and order.}
\label{tab:exper_accuracy_schrod_3}
\begin{tabular}{|c|c|c|c|c|c|c|}
\hline
Parameters & Scheme & N = 10 & N = 20 & N = 50 & N = 100 & order \\  \hline
$a = 1, b = 1, \omega = 1$ & (\ref{divergent_1}) & 2.18+1 & 6.30+0 & 1.05+0 & 2.62-1 & 1.94 \\  \hline
$a = 1, b = 1, \omega = 1$ & (\ref{comp_scheme}) & 6.79-1 & 4.88-2 & 1.27-3 & 8.05-5 & 3.94 \\  \hline

$a = 1, b = 2, \omega = 2$ & (\ref{divergent_1}) & 2.60+4 & 8.50+3 & 1.48+3 & 3.72+2 & 1.89 \\  \hline
$a = 1, b = 2, \omega = 2$ & (\ref{comp_scheme}) & 5.02+3 & 3.77+2 & 1.01+1 & 6.38-1 & 3.93 \\  \hline

$a = 1, b = 0.1, \omega = 1$ & (\ref{divergent_1}) & 9.79-2 & 2.31-2 & 3.91-3 & 9.61-4 & 2.03 \\  \hline
$a = 1, b = 0.1, \omega = 1$ & (\ref{comp_scheme}) & 4.75-4 & 3.39-5 & 8.48-7 & 5.52-8 & 3.96 \\  \hline

$a = 1, b = 2, \omega = 10$ & (\ref{divergent_1}) & 5.67+4 & 1.71+4 & 2.88+3 & 7.39+2 & 1.90 \\  \hline
$a = 1, b = 2, \omega = 10$ & (\ref{comp_scheme}) & 1.22+4 & 7.94+2 & 2.28+1 & 1.39+0 & 3.96 \\  \hline
\end{tabular}
\end{table}

All eigenvalues of the transition operator $M=-A^{-1}_{new}A_{old}$ are unimodal:
\[
|\lambda_k|=1,\quad k=1,\ldots,N-1
\]
(see Fig.~\ref{fig:eigens}) and the matrix $M$ has no non-trivial Jordan blocks. Therefore, in the standard Euclidean space $\mathbb{C}^{N-1}$ there is a positive definite quadratic form which is conserved according to the finite difference equation 
(\ref{ll_comp_scheme}). However, 
it is not trapezoidal or Simpson quadrature on the segment $x\in [0,\,2\pi]$ of $|\Psi(n\tau,\,x)|^2$.
The coefficients of the quadrature are not constants with respect to the index $n$.

\begin{figure}[h!]
\begin{center}
	\includegraphics[scale=.65]{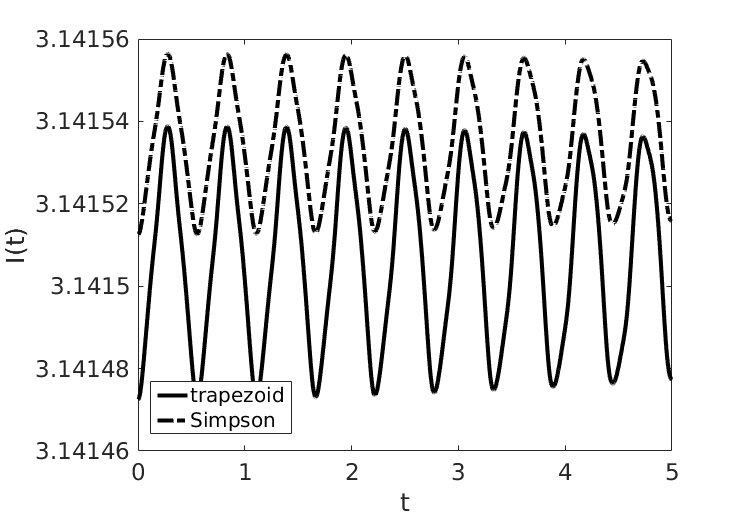}
	\caption{
First integral of Leontovich -– Levin equation (\ref{Schro_eq}) $I(t) = \int\limits_0^{2\pi} |\Psi (t,\,x)|^2\,dx$ 
with $\Psi(0, x) = sin(x); \theta(x) = i [cos^2(x)+1]; f \equiv 0$, computed numerically using trapezoidal 
and parabolic (Simpson) quadrature. $N = 50, |\nu| = 1$. We checked numerically that the amplitude of oscillations at small step $h$ is proportional 
to $h^3$ for both quadratures. 
	}
	\label{fig:first_int}
\end{center}
\end{figure}

There are oscillations of these quadratures (see Fig.~\ref{fig:first_int}). However, the amplitude of the oscillations
decreases as $O(N^3)$ at $N\to \infty$.

Thus, there is a positive definite quadratic form which conserved according to the
finite difference equation (\ref{ll_comp_scheme}) and tends to the standard Euclidean metric as $N \to \infty$.
The length of the arc with eigenvalues of $M$ (see Fig. \ref{fig:eigens}) depends on the Courant parameter 
$|\nu|$. If the spatial step $h$ is fixed, and the temporal step $\tau$ tends to zero, the arc contracts 
to $\lambda=1$. As about the alternative limit $\tau \to \infty$, the arc develops on the whole segment 
$[\pi, 2\pi]$.

We consider the spectrum of the Dirichlet problem for (\ref{Schro_eq}). $Spec~M_{Dirichlet}$ of the operator $M_{Dirichlet}$ on the subspace of grid functions $u$ such that $u_0 = u_N = 0$. According to our numerical experiments, $Spec~M_{Dirichlet}$ is situated at the arc of the unit circle on the complex plane, see Fig.~\ref{fig:eigens}. The left boundary of the arc depends on $\nu$ and $N$: if $\nu$ is fixed and $N \to \infty$, then the angle $\alpha \to const$. If $\nu \to 0$, then $ \alpha \to 0$; 
if $\nu \to \infty$, then $\alpha \to \pi$. The spectrum $Spec~M_{Neumann}$
 for the Neumann problem is wider, see additional "star" on the Fig.~\ref{fig:eigens}.
 
\pagebreak
\section{Summary and discussion}

We have presented the $4$-th order compact implicit scheme which approximates mixed problems for the 1D parabolic equation with a variable coefficient and for the Leontovich -- Levin equation. We have confirmed the stability and convergence of the scheme by various numerical experiments, and it is the main result of the paper. We studied the spectral structure of transition operators of the scheme to explain the results. The scheme conserves the first integral for the homogeneous Leontovich -- Levin equation.

We have compared the scheme with the classic implicit scheme and the advantages of the new scheme are clear. The number of arithmetic operations for both considered schemes is similar. 

This approach may be used for approximation of various linear PDEs with variable coefficients. Moreover, it can be developed for the approximation of weakly non-linear PDE like the non-linear 
Schr\"odinger equation or the Fisher -- Kolmogorov -- Petrovsky -- Piskunov equation. We are going
to describe the extension in another article.

We have considered here the Dirichlet and Neumann boundary conditions. The compact scheme is sensitive to the quality of the Neumann condition’s approximation. The special compact approximation of the Neumann condition was constructed. The function $f$ and its derivatives must be included into the difference boundary conditions to avoid the loss of order. The compact approach to the boundary condition’s approximation may be developed for other types of boundary conditions.
The compact schemes are preferable for many-dimensional problems, too. Iteration approaches are effective for implementation of the implicit schemes.


\section{Acknowledgements}

The article was prepared within the framework of the Academic Fund Program at the National Research 
University Higher School of Economics (HSE) in 2016--2017 and 2018--2019 (grants \# 16-05-0069 and 18-05-0011) and supported within 
the framework of a subsidy granted to the HSE by the Government of the Russian Federation for the 
implementation of the Global Competitiveness Program.
\\
The authors gratefully acknowledge the anonymous referee whose comments were helpful to improve the initial version of the paper.

\bibliographystyle{ieeetr}
\clearpage

\end{document}